\documentclass[aps,pre,twocolumn,superscriptaddress,showkeys]{revtex4-1}
\usepackage{natbib}
\usepackage{amsmath,amssymb}
\usepackage{graphicx}
\usepackage{subfigure}
\usepackage{dcolumn}
\usepackage[latin1]{inputenc}
\usepackage{float}
\usepackage{color}
\usepackage{amsmath}
\definecolor{rojo}{rgb}{1,0,0}
\definecolor{verde}{rgb}{0,0.8,0.2}
\definecolor{azul}{rgb}{0,0,1}
\definecolor{rosa}{cmyk}{0,1,0,0}
\usepackage[toc,page]{appendix} 
\usepackage{latexsym}
\usepackage{amsmath}
\usepackage{amsthm}
\usepackage{amsfonts}
\usepackage{amssymb}
\usepackage{color}
\usepackage{epsfig}
\usepackage{hyperref}
\hypersetup{colorlinks=true, citecolor=blue}
\usepackage{soul}

\usepackage{braket}
\usepackage{mathtools}

\newcommand{\depa}[2]{\frac{\partial #1}{\partial #2}}

\setstcolor{red}

\begin{document}

\title{{Proximity induced spin-orbit effects in graphene on Au}}

\author{Alejandro Lopez}
\affiliation{Centro de F\'isica, Instituto Venezolano de Investigaciones Cient\'ificas (IVIC), 
Apartado 21827, Caracas 1020 A, Venezuela }%

\author{Luis Colmenarez}
\affiliation{Centro de F\'isica, Instituto Venezolano de Investigaciones Cient\'ificas (IVIC), 
Apartado 21827, Caracas 1020 A, Venezuela }%

\author{Mayra Peralta}
\email{mperalta@yachaytech.edu.ec}
\affiliation{Yachay Tech, School of Physical Sciences \& Nanotechnology, 100119-Urcuqu\'i, Ecuador}
\affiliation{{Departamento de F\'isica,} Centro de Nanociencias y Nanotecnolog\'ia, Universidad Nacional Aut\'onoma de M\'exico,  Apdo. Postal 14, C.P. 22800, Ensenada B.C., M\'exico.}

\author{Francisco Mireles}
\affiliation{{Departamento de F\'isica,} Centro de Nanociencias y Nanotecnolog\'ia, Universidad Nacional Aut\'onoma de M\'exico,  Apdo. Postal 14, C.P. 22800, Ensenada B.C., M\'exico.}

\author{Ernesto Medina}
\email{emedina@yachaytech.edu.ec}
\affiliation{Yachay Tech, School of Physical Sciences \& Nanotechnology, 100119-Urcuqu\'i, Ecuador}
\affiliation{Centro de F\'isica, Instituto Venezolano de Investigaciones Cient\'ificas (IVIC), 
Apartado 21827, Caracas 1020 A, Venezuela }%

\date{\today}

\begin{abstract}
We introduce a $p_z$-$d$ coupling model Hamiltonian for the  $\pi$-graphene/Au  bands that predicts a rather large intrinsic  spin-orbit (SO) coupling  as are being reported in recent experiments and DFT studies. Working within the analytical Slater-Koster tight-binding approach we were able to identify the overlapping orbitals of relevance in the enhancement of the SO coupling for both, the sublattice symmetric (BC), and  the ATOP (AC) stacking configurations. Our model
effective Hamiltonian reproduces quite well the experimental spectrum for the two registries, and in addition, its shows that the hollow site configuration  (BC),  in which the A/B sites remain symmetric, yields the larger increase of the SO coupling.  We also explore the Au-diluted case keeping the BC configuration and showed that it renders the preservation of the SO-gap with a similar SO interaction enhancement as the undiluted case but with a smaller graphene-gold distance.

\end{abstract}

\maketitle

\section{Introduction}
Graphene/transition metal interfaces have been recognized as a very attractive hybrid systems \cite{MacDonald,Tok,Wehling,Saha,Avsar} since its proximity may lend fascinating properties that the otherwise isolated graphene layer lacks. Namely, the gate controlled dopability \cite{Castro}, the transfer of ferro and antiferromagnetism \cite{MacDonald,Peralta}, and the induced strong spin-orbit coupling \cite{Marchenko}, just to mention a few. In addition, more complex transition metal substrates such as transition metal dichalcogenides on few-layer graphene hybrid spin-valves  have shown to induce opto-valley spin-injection due to its large SO coupling \cite{MitraFabian,LuoOptoSpin}. Such graphene-hybrid materials are of much current interest because they can provide unique spintronics applications.

All the changes on the physical properties in graphene mentioned above are operated by {\it proximity effects}. Such effects have the additional advantage of interfering only weakly with graphene's mobility in contrast for example, with the adatom/doping approaches that locally deform the lattice \cite{Avsar}, in order to enhance the spin orbit coupling \cite{Balakrishnan,Tok,Brey}. The intrinsic spin-orbit coupling  in graphene generates a gap at the $K$-Dirac point close to 20$\mu$eV\cite{Rashba}. One can induce an additional source of SO interaction {\it i.e.} the Rashba SO coupling (RSO), as long as the space inversion symmetry of graphene is broken by a substrate, the presence of an external electric field or adatoms \cite{Gmitra}. The spin splitting of doubly degenerate pristine graphene energy bands is a signature of this interaction. The {RSO} coupling arising due to typical external electric fields (gating) is estimated to be rather weak, of the order of just $5 \mu$eV for a field of 1 V/nm \cite{Rashba}. As SO and {RSO} couplings are responsible for interesting quantum phases like the spin Hall effect \cite{Balakrishnan}, topological quantum spin Hall effect \cite{Kane} and spin active persistent currents\cite{BolivarMedinaBerche}, their small magnitude in pristine graphene makes these interesting features unobservable. Many efforts have then been directed to enhance the spin-orbit interaction.

Here we can mention three experimentally accesible ways to enhance the SO strength in graphene: {\it  i}) Adding light adatom impurities, such as hydrogen, introduces local warpings of graphene and increases the SO coupling to about $10$ meV \cite{Castro2}, close to the atomic SO strength of carbon. However this approach has the disadvantage of the concomitant reduction of the electron mobility \cite{Drexler,Zhang}. {\it  ii}) Bending the graphene sheet and producing tubes, cones and Bucky balls with additional defects. This geometrical variation also produces a SO strength of the order of $10$ meV\cite{AndoSONanotubes,HuertasGuinea,KonschuhFabian},  and  {\it  iii}) Placing the graphene layer in contact with heavy atoms to induce SO proximity effects whiles interfers weakly with the mobility of graphene. For instance, a few-layer semiconducting tungsten disulphide increases the SO interaction on a single layer graphene to about $17$ meV \cite{Avsar}. First principles calculations show that BiFeO$_3$ induces a {RSO} on graphene of approximately $1.26$ meV \cite{Zhenhua}. Also, a strong spin orbit splitting has been measured in Graphene/Pb/Ir(111) of about $30$meV \cite{Otrokov}.

Recently, Marchenko {\it et al.} \cite{Marchenko} attempted the third approach by building a Au-graphene interface, generating a large spin-orbit splitting of $\sim100$ meV on graphene according to its spin-resolved photoemission measurements. The resulting band structure and SO-gap magnitudes were strongly dependent on the gold atom-graphene stacking, finding that the sublattice symmetric BC stacking -- as opposed to the AC (ATOP) configuration -- is very effective in enhancing the SO coupling. The authors also report that the epitaxial graphene develops a giant spin-orbit gap of about $\sim$70 meV and managed to fit the modifications to the pristine graphene band structure by adjusting the parameters of the Kene-Mele Hamiltonian\cite{Kane}. The system of Graphene/Au/Ni(111) was later analyzed in a very recent study, in which the giant {RSO} is explained as a nanoscale effect, using a detailed STM characterization and DFT modeling \cite{Krivenkov}. In this study, the authors find that the giant {RSO} generated in Graphene/Au/Ni(111), is due to a decreased gold-graphene equilibrium distance as a result of a graphene-Ni interaction.

In this paper we propose an analytical Slater-Koster model for Au-graphene interface assuming that the gold atoms are either above of every hollow position of the graphene lattice (BC or HCP stacking), or at the same hollow site in a diluted configuration. Such registries, according to experiments, are the most effective raising the strength of the SO coupling. We consider the overlaps between the graphene $p_z$ orbitals and the $5d$ orbitals of Au. Using lowest order perturbation theory, we arrive at an effective Hamiltonian which includes the SO coupling inherited from the Au by proximity effects, corrections to the Fermi velocity and the resulting net doping\cite{Khomyakov} of graphene. Our analytical model reproduces the spectrum derived by DFT computations and the experimental results of ref.\cite{Marchenko} for both the {AC stacking (or ATOP)} and the hollow site configurations.

\section{Analytical Slater-Koster model graphene on Au \label{analytical}}

We will now develop a perturbative model to lowest order to derive analytical expressions that describe how the spin-orbit (SO) coupling is inherited from the gold surface atoms to the graphene. The SO strength is derived from the intrinsic atomic SO coupling of gold, and we derive explicit dependencies of the probability amplitude to hop between the $p_z$ orbital perpendicular
to the graphene plane through the paths that connect these orbitals through the gold surface $5d$ orbitals {(other models that include d orbitals for predicting the spin orbit coupling of graphene can be found in refs. \cite{KonschuhFabian,Boykin,BoykinNayak})}.

\begin{figure}[h]
	\centering
	\includegraphics[width=8.0cm,height=10.5cm]{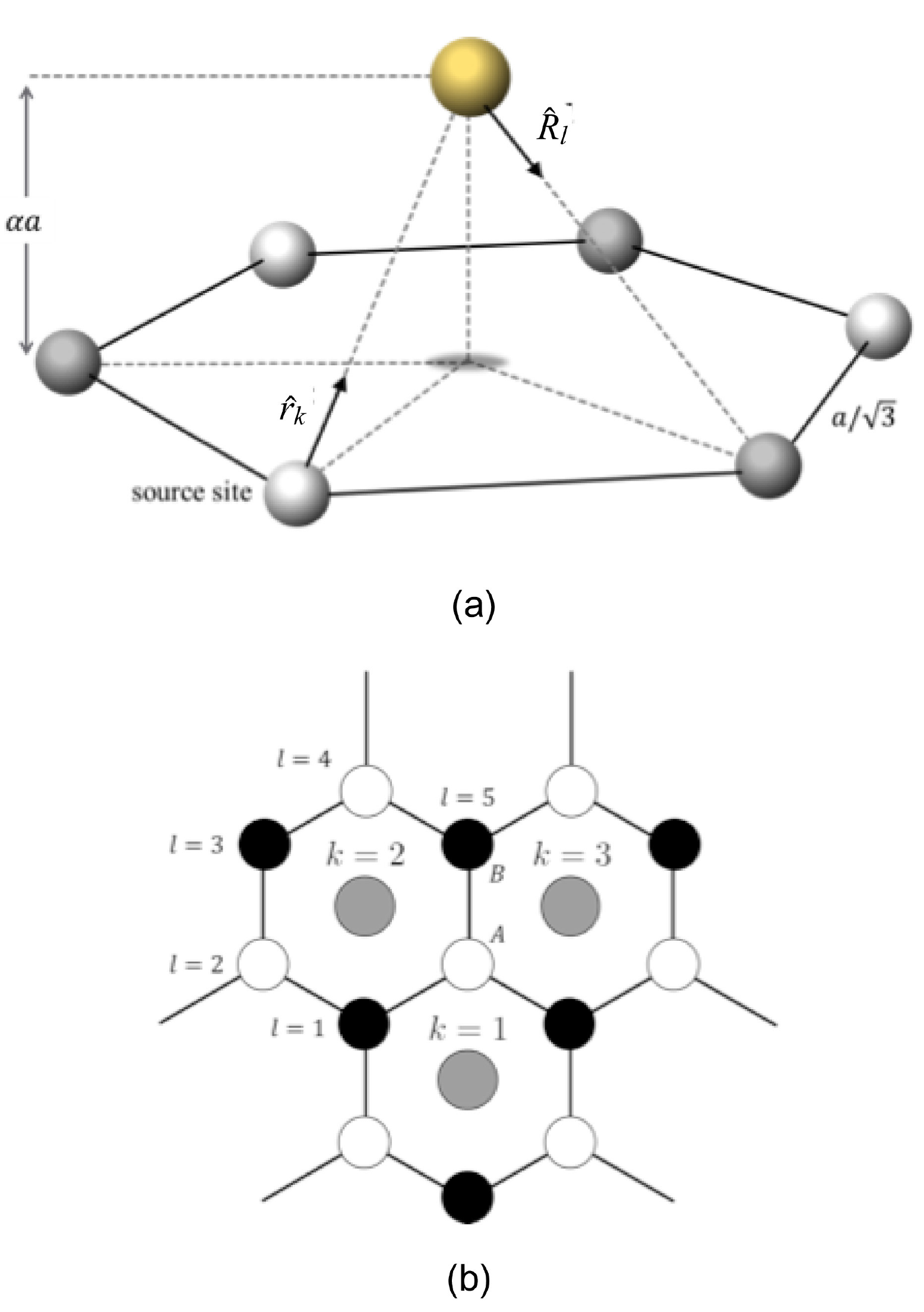}
	\caption{  {(a)}: unitary vector {$\hat{r}_k$} that points from the carbon site to the gold atom, and {$\hat{R}_l$} that points back to graphene sites, $\alpha a$ is the graphene-gold lattice distance, $a$ is the graphene's lattice parameter, therefore, every carbon atom is separated $a/\sqrt{3}$ from its nearest neighbors. {(b)}: labels of the neighbors involved in the interaction with the source site in graphene (white site at the center). The white (black) sites  belongs to A (B) sub-lattice in graphene and they are labeled by $l$, gray circles represent sites in gold and they are labeled by $k$. The $l$ labels for $k=2$ are the only ones drawn, $l=6$ will be always the source site for each $k$. \label{etiquetas1}}
\end{figure}

The system we are studying is composed by graphene on top of a gold monolayer in the HCP configuration, namely, the layer of graphene is located in such a way that every atom of gold is positioned at the graphene's hollows. In this configuration, electrons can move from one $p_z$-orbital to another either as in pristine graphene or by using the available $5d$-orbitals of gold as a bridge between them. We set the origin of coordinates at the electron source site {(central atom A of Figure \ref{etiquetas1}(b))} and describe the positions of the three neighbors in gold by $\hat{r}_{k}$, $k=1,2,3$, as we show in the Figure \ref{etiquetas1}. The positions of the neighbors {in the graphene lattice}, counting up to third nearest neighbors from the source site, are described by $\hat{R}_{l}$, $l=1,\dots,6$, and their origins are at the corresponding $k$ sites. The vectors are written as
{
\begin{eqnarray}
	\begin{split}
	\hat{r}_{k} = (n_{kx}, n_{ky}, n_{kz}), \\
	\hat{R}_{l} = (n_{lx}, n_{ly}, n_{lz}).
	\end{split}
\end{eqnarray}
}

For example, the two vectors drawn in the upper panel of Figure \ref{etiquetas1}, are written as
{
\begin{eqnarray}
	\begin{split}
	\hat{r}_{k=2} &=& \frac{1}{\zeta}\left(-\frac{1}{2},\frac{1}{2\sqrt{3}},\alpha\right), \\
	\hat{R}_{l=5} &=& \frac{1}{\zeta}\left(\frac{1}{2},\frac{1}{2\sqrt{3}},-\alpha\right),
	\end{split}
\end{eqnarray}}
where $\alpha$ is the ratio between the perpendicular graphene-gold distance and the lattice parameter  $a=2.46$\AA $\,$, and $\zeta=\sqrt{1/3+\alpha^2}$ is the ratio between the distance from the carbon site to the gold atom (see Figure 1) and $a$.  The perpendicular distance $\alpha a$ from the graphene plane to the site in gold is estimated around 2.5\AA $\,$ by {\it ab initio} calculations \cite{Marchenko}, therefore, $\alpha\approx 1.02$ and $\zeta \approx 1.17$. Note also that $n_{kz}=-n_{lz}=\alpha/\zeta\approx0.87$ for all $k$ and $l$.

\subsection{Hopping Integrals \label{HP}}

 {In this section we write the hopping integrals, $E^k_{\mu,\mu'}$, between orbitals $\mu$ and $\mu'$ at site $k$ as a linear combination of Slater-Koster parameters, $V_{\mu,\mu',\pi(\sigma)}$, in the notation used in \cite{SlaterKoster}}. The $E^{k}_{\mu,\mu'}$ terms, relate to $p_z$ (graphene carbon) and $d$ (gold) orbitals between which the electrons hop from graphene to gold and back. Omitting the site index for clarity, we have
 \begin{equation}
 E_{z,xz} = n_xV_{p_z d_{xz}} = E_{xz,z},
 \end{equation}
 {where $V_{p_z d_{xz}}\equiv\sqrt{3}n_z^2V_{pd\sigma} + \left(1-2n_z^2\right)V_{pd\pi}$}, and $n_x$ is the $x$ component
 of the unit vector that connects the orbitals $p_z$ at the source site and the $d_{xz}$ orbital at site $k$.
 \begin{eqnarray}
 E_{z,yz} &=& n_y V_{p_z d_{xz}} = E_{yz,z} , \\
 E_{z,z^2} &=& V_{p_z d_{z^2}} = E_{z^2,z},
 \end{eqnarray}
 where {$V_{p_z d_{z^2}} \equiv \sqrt{3}n_z(n_x^2+n_y^2)V_{pd\pi} - \frac{1}{2}n_z(n_x^2 + n_y^2 - 2n_z^2)V_{pd\sigma}$},
 {
 \begin{eqnarray}
 E_{z,x^2-y^2} &=& (n_x^2-n_y^2) V_{p_z d_{x^2-y^2}} = E_{x^2-y^2,z} ,
 \end{eqnarray}}
 where {$V_{p_z d_{x^2-y^2}} \equiv n_z\left(\frac{\sqrt{3}}{2}V_{pd\sigma} - V_{pd\pi}\right)$}, and
 \begin{eqnarray}
 E_{z,xy} &=& n_x n_y V_{p_z d_{xy}} = E_{xy,z},
 \end{eqnarray}
 where {$V_{p_z d_{xy}} \equiv n_z(\sqrt{3}V_{pd\sigma} - 2V_{pd\pi}) = 2V_{p_z d_{x^2-y^2}}$}. The complex conjugated integrals are given by $\braket{\mu|H|\mu'} = (-1)^{l+l'}\Braket{\mu'|H|\mu}$, with $l(l')$ the quantum number of the angular momentum associated with the $\mu(\mu')$ orbital. 
 
 We summarize the results of the hopping integrals, for hops from gold to graphene, in the Table \ref{sk_matrix}, where we use the definition:
{
 \begin{equation}
 	(u_x, u_y) \equiv \left(\frac{1}{2}(n_x^2-n_y^2),n_x n_y\right).
 \end{equation}}
\begin{table}[h!]
	\caption{Hopping Integrals between $p_z$  ($k$ from source site to gold) and $d$ ($l$ from gold back to graphene) orbitals {(see \cite{SlaterKoster})}} \label{sk_matrix}
	\begin{equation*}
	\begin{array}{ l  c  c  c  c }
		\toprule
		\mu         		& E_{z,\mu}^k      		& E_{\mu, z}^k     		& E_{\mu,z}^l         		& E_{z,\mu}^l           \\ \hline
		d_{z^2}     	& V_{pd_{z^2}}     		& V_{pd_{z^2}}     		& V_{pd_{z^2}}			& V_{pd_{z^2}}          \\
		d_{xz}      		& n_{kx} V_{pd_{xz}} 	& n_{kx} V_{pd_{xz}} 	& -n_{lx} V_{pd_{xz}} 	& -n_{lx} V_{pd_{xz}} \\
		d_{yz}      		& n_{ky} V_{pd_{xz}} 	& n_{ky} V_{pd_{xz}} 	& -n_{ly} V_{pd_{xz}} 	& -n_{ly} V_{pd_{xz}} \\
		d_{x^2-y^2} 	& u_{kx} V_{pd_{xy}} 	& u_{kx} V_{pd_{xy}} 	& u_{lx} V_{pd_{xy}}  	& u_{lx} V_{pd_{xy}}  \\
		d_{xy}      		& u_{ky} V_{pd_{xy}} 	& u_{ky} V_{pd_{xy}} 	& u_{ly} V_{pd_{xy}}  	& u_{ly} V_{pd_{xy}}  \\ \toprule
	\end{array}
	\end{equation*}
\end{table}

\subsection{Spin-Orbit Coupling Matrix}
 The spin-orbit term, $H_{SO}$ is the following:
\begin{eqnarray}
	H_{SO} &=& \frac{e}{2m_0^2c^2}(\nabla V\times \mathbf{p})\cdot\mathbf{S}\nonumber\\
	             &=& \frac{\lambda}{2}\left[L_+S_- + L_-S_+ + 2L_zS_z\right]
\end{eqnarray}
where $e$ the electron charge, $m_0$ is the free electron mass, $c$ the speed of light, $V$ the atomic electric potential,  $\bm p$ is the momentum operator, and   ${\bf S}=\frac{\hbar}{2}\bm\sigma$  the spin vector operator, with $\bm\sigma$ the vector of the spin Pauli matrices.  Here $\lambda=\frac{1}{r}\depa{V}{r}\frac{e}{2m_0^2c^2}$, and the operators $L_\pm=L_x\pm iL_y$,  and $S_\pm=S_x\pm iS_y$, with $L_i$ and  $S_i$ are the {\it i=\{x,y\}} components of the angular momentum and spin operators, respectively. The relevant orbitals involved for the gold atoms are the $5d$ orbitals, which can be written as a linear combination of the spherical harmonics as: 
\begin{eqnarray}
	\begin{split}
	\Ket{d_{xy}} &= \frac{i}{\sqrt{2}}\left(\Ket{2,-2} - \Ket{2,2}\right), \\
	\Ket{d_{yz}} &= \frac{i}{\sqrt{2}}\left(\Ket{2,-1} + \Ket{2,1}\right), \\
	\Ket{d_{z^2}} &= \Ket{2,0}, \\
	\Ket{d_{xz}} &=\frac{1}{\sqrt{2}}\left(\Ket{2,-1} - \Ket{2,1}\right), \\
	\Ket{d_{x^2-y^2}} &= \frac{1}{\sqrt{2}}\left(\Ket{2,-2} + \Ket{2,2}\right).
	\end{split}
\end{eqnarray}

{ A detailed derivation, for the spin orbit coupling term used in tight binding, can be found in \cite{KonschuhFabian,Boykin,BoykinNayak,Yao}. The SO couplings between $d$ orbitals are presented in \cite{KonschuhFabian}.}

\subsection{$p_z-p_z$ Coupling through $d$ Orbitals}

Now, we suppose that the eigenfunctions are expanded over the atomic orbital basis as 
\begin{equation}
	\ket{\Psi} = \sum_{\mu,\alpha} g_{\mu,\alpha} a^\dagger_{\mu,\alpha}\ket{0},
\end{equation}
where $\mu=\{s, p_x, p_y, p_z, d_{xy},\dots\}$ represents all possible orbitals, $\alpha$ is the orbital position, $g_{\mu,\alpha}$ is the expansion coefficient and $a^\dagger_{\mu,\alpha}$ is the corresponding creation operator. In general, the matrix element resulting from a coupling between an initial state $\ket{\mu, \alpha}$ and the general state $\ket{\Psi}$ is
\begin{equation}
	\braket{\mu,\alpha|\hat{H}|\Psi} = \braket{\mu,\alpha|\epsilon|\Psi} = \epsilon g_{\mu,\alpha},
\end{equation}
where $\epsilon$ is the eigenvalue of the full Hamiltonian involving all the couplings present.

We choose the orbital $p_z$ perpendicular to the graphene plane ($\mu=z$, for simplicity) at position $\alpha=0$ as one mobile electron bearing orbital, $\ket{p_z(0)}$, so the coupling between the source site and its six neighbors (three in graphene and three in gold)  is given by the equation{
\begin{eqnarray}
	(\epsilon - \epsilon_p) g_{z} &=& V_{pp\pi} \sum_{j=1}^{3}g_{z,j} + \sum_{k=1}^{3}\sum_{\mu}E_{z,\mu}^k c_{\mu, k}, \label{eq:a_z}
\end{eqnarray}}
where $\epsilon_p$ is the bare energy of the electron at the $p_z$ orbital. The first term on the right hand side represents the connection with the three nearest neighbors in graphene, and the second term with the tree nearest neighbors on the gold surface. {$V_{pp\pi}$} is the Slater-Koster parameter that represent the energy to go from site A to B by a $\pi$ overlap of the $p_z$ orbitals. The $E_{z,\mu}^k$ are the hopping integrals between $p_z$ orbital of graphene and $\mu=\{d_{xz}, d_{yz}, d_{z^2}, d_{x^2-y^2}, d_{xy}\}$ orbitals of gold, where the $d$ orbital is at site $k$. Note that we labeled the expansion coefficients $g$ and $c$ to distinguish interactions with sites on graphene or gold, respectively.

For writing the hopping integrals, we use Table \ref{sk_matrix} and the results of the previous subsection. For example if we choose the $d_{z^2}$ orbital ($\mu=z^2$, for simpli\-city) and the positions $\alpha=k=\{1,2,3\}$ as the initial state, $\ket{d_{z^2}(k)}$, the hopping integrals between sites in gold and its six neighbors in graphene are:
{
\begin{eqnarray}
(\epsilon-\epsilon_d) c_{z^2,k}	&=&  -i\sqrt{3}s_y\xi_dc_{xz,k} + i\sqrt{3}s_x\xi_dc_{yz,k} \nonumber\\
								& & + \sum_{l}E_{z^2,z}^{l}g_{z,l}\nonumber\\
						&=&  -i\sqrt{3}s_y\xi_dc_{xz,k} + i\sqrt{3}s_x\xi_dc_{yz,k} \nonumber\\
								& & + V_{pd_{z^2}} \sum_{l}g_{z,l}. \label{cz2_eq}
\end{eqnarray}}

Analogously, for the rest of $d$ orbitals with $\mu=\{xz, yz, x^2-y^2, xy\}$, (Se Appendix A Eqs. \ref{hintegrals1}-\ref{hintegrals5}). 

Substituting the coefficients $c_{\mu,k}$, shown in Appendix A (Eqs. \ref{ccoef1}-\ref{ccoef5}), into Eq.(\ref{eq:a_z}) and, after some algebra, the resulting equation for the graphene's coefficients, $g_{z,l}$, can be written in the following form:
{
{\footnotesize
\begin{eqnarray}
	\tilde{\epsilon} g_{z}	&=&  V_{pp\pi} \sum_{j}g_{z,j} \nonumber\\
					& & + \frac{1}{\epsilon_p-\epsilon_d}\sum_{k,l}\left[V_{pd_{z^2}}^2 - V_{pd_{xz}}^2(\hat{r}_k\cdot\hat{R}_l + n_z^2) + V_{pd_{xy}}^2\right] g_{z,l}\nonumber\\
					& & + i\frac{\sqrt{3}\xi_dV_{pd_{xz}}V_{pd_{z^2}}}{(\epsilon_p-\epsilon_d)^2} \sum_{k,l}\left[\vec{s}\times(\hat{r}_k - \hat{R}_l)\right]_z g_{z,l}\nonumber\\
					& & + i\frac{\xi_d V_{pd_{xz}}^2}{(\epsilon_p-\epsilon_d)^2} s_z\sum_{k,l}(\hat{r}_k\times\hat{R}_l)_z g_{z,l}\nonumber\\
					& & - i\frac{2\xi_dV_{pd_{xy}}^2}{(\epsilon_p-\epsilon_d)^2}s_z \sum_{k,l}(\vec{u}_{\, k}\times\vec{u}_{\, l})_z g_{z,l}\nonumber\\
					& & + i\frac{\xi_dV_{pd_{xz}}V_{pd_{xy}}}{(\epsilon_p-\epsilon_d)^2} \sum_{k,l}\left[s_x\left(\hat{r}_k\times\vec{u}_{\, l} + \hat{R}_l\times\vec{u}_{\, k}\right)_z \right. \nonumber\\
					& & \left.\qquad\qquad\qquad\qquad- s_y\left(\hat{r}_k\cdot\vec{u}_{\, l} + \hat{R}_l\cdot\vec{u}_{\, k}\right)\right] g_{z,l},
\label{HamiltonianIdentify}
\end{eqnarray}}}
where $\tilde{\epsilon}= \epsilon - \epsilon_p - \xi_{sz}^2/(\epsilon_p-\epsilon_s)$ is the perturbed energy parameter. 
We have eliminated from the equations the coefficients associated with gold ($c$) and only the graphene coefficients {($g$)} appear. Thus we
have renormalized the couplings between the graphene orbitals, $p_z$, through the couplings to gold.

In the order they appear above, the terms represent: (1) the kinetic energy of pristine graphene, (2) the kinetic energy of an electron that uses gold sites to bridge between the source site and its neighbors up to second nearest neighbors, note that all $d$ orbitals are taken into account; (3) the intrinsic SO term built from one of the following four paths
\begin{equation*}
	\begin{array}{c c c c c c c}
		p_z & \xrightarrow{V_{pd_{xz}}}  & d_{xz} || d_{yz} & \xrightarrow{SO} & d_{z^2}          & \xrightarrow{V_{pd_{z^2}}} & p_z , \\
		p_z & \xrightarrow{V_{pd_{z^2}}} & d_{z^2}          & \xrightarrow{SO} & d_{xz} || d_{yz} & \xrightarrow{V_{pd_{xz}}}  & p_z .
	\end{array}
\end{equation*}
Note that the coupling between $d_{yz}$ and $p_z$ is expressed in terms of $V_{pd_{xz}}$ as can be seen from the relations in section \ref{HP}.
(4) the intrinsic SO term in which the electron follows one of the following two paths
\begin{equation*}
	\begin{array}{c c c c c c c}
		p_z & \xrightarrow{V_{pd_{xz}}} & d_{xz} & \xrightarrow{SO} & d_{yz} & \xrightarrow{V_{pd_{xz}}} & p_z , \\
		p_z & \xrightarrow{V_{pd_{xz}}} & d_{yz} & \xrightarrow{SO} & d_{xz} & \xrightarrow{V_{pd_{xz}}} & p_z ;
	\end{array}
\end{equation*}
(5) the intrinsic SO term in which the electron follows one of the paths
\begin{equation*}
	\begin{array}{c c c c c c c}
		p_z & \xrightarrow{V_{pd_{xy}}} & d_{xy}      & \xrightarrow{SO} & d_{x^2-y^2} & \xrightarrow{V_{pd_{xy}}} & p_z , \\
		p_z & \xrightarrow{V_{pd_{xy}}} & d_{x^2-y^2} & \xrightarrow{SO} & d_{xy}      & \xrightarrow{V_{pd_{xy}}} & p_z ;
	\end{array}
\end{equation*}
and finally, (6) the intrinsic SO term in which the electron follows one of the following eight paths
\begin{equation*}
	\begin{array}{c c c c c c c}
		p_z & \xrightarrow{V_{pd_{xz}}} & d_{xz} || d_{yz}      & \xrightarrow{SO} & d_{xy} || d_{x^2-y^2} & \xrightarrow{V_{pd_{xy}}} & p_z , \\
		p_z & \xrightarrow{V_{pd_{xy}}} & d_{xy} || d_{x^2-y^2} & \xrightarrow{SO} & d_{xz} || d_{yz}      & \xrightarrow{V_{pd_{xz}}} & p_z,
	\end{array}
\end{equation*}
where we used the symbol $||$ as the logical OR operator. Each diagram goes from the source site ($p_z$ orbital) to a $k$ site in gold ($d$-orbitals), where the SO coupling of the gold is involved, and back to an $l$ site in graphene ($p_z$ orbital). In the next subsections we obtain simplified expressions for all the previous terms.

\subsubsection{Kinetic Term $H_K$}

\begin{figure}[h]
	\centering
	\includegraphics[width=.3\textwidth]{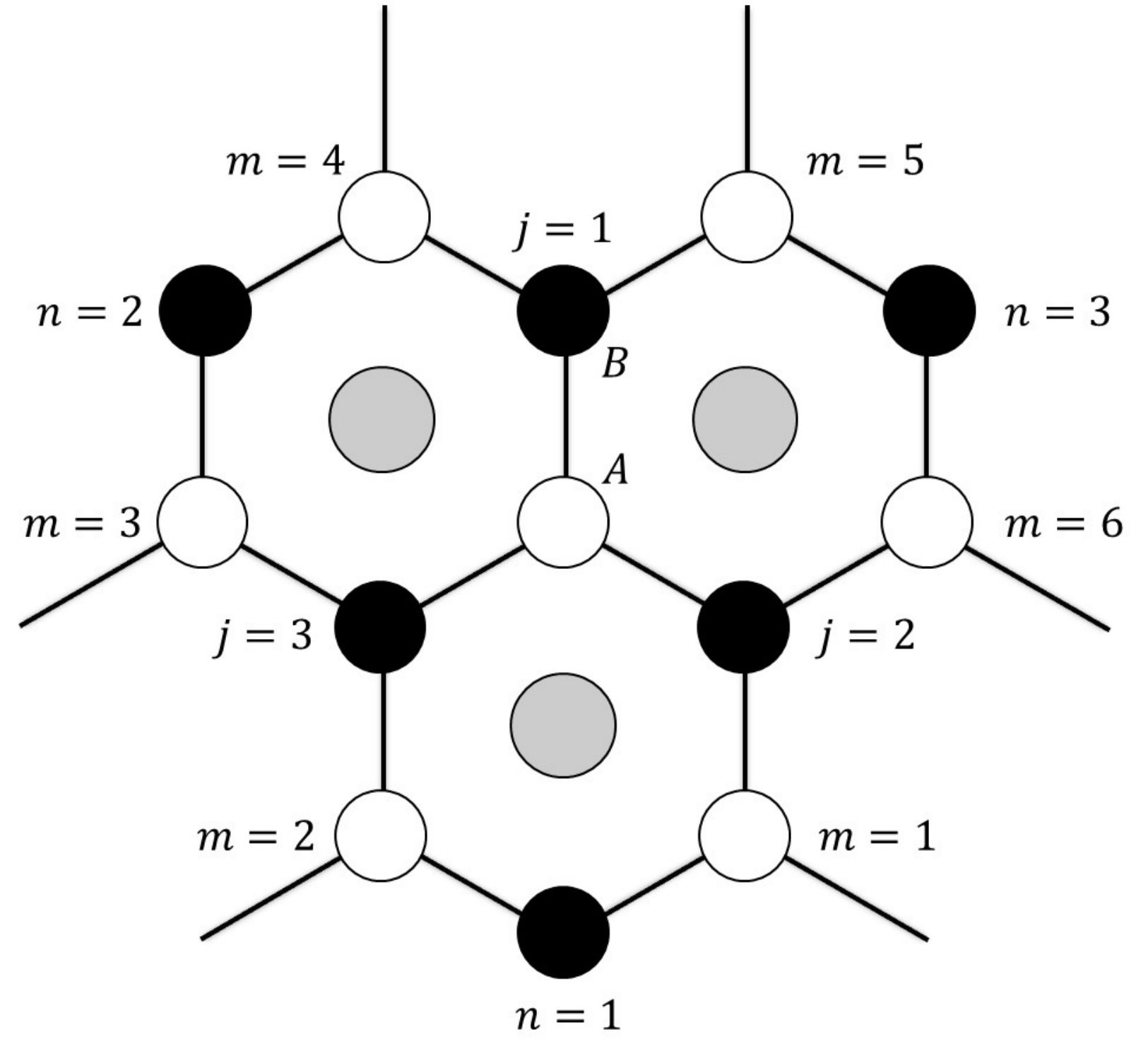}
	\caption{Labels of the all neighbors involved in the interaction with the source site in graphene (white site at the center). The white (black) sites  belongs to A (B) sub-lattice in graphene and they are labeled by $j$ for nearest neighbors, $m$ second neighbors and $n$ third neighbors, and gray circles represent gold sites. \label{etiquetas2}}
\end{figure}

Having identified the different contributions to the Hamiltonian from Eq. \ref{HamiltonianIdentify} we rewrite the Kinetic term contributions as
\begin{eqnarray}
	H_K &=& t_1\sum_{j=1}^3b_{z,j} + t_2\sum_{m=1}^6a_{z,m} + t_3\sum_{n=1}^3b'_{z,n}
\end{eqnarray}
where we have labeled the expansion coefficients as $b$, $a$ and $b'$ and identified with $j$ the first, with $m$ the second and $n$ the third nearest neighbors from the source site, respectively (see Figure \ref{etiquetas2}). This notation leads to a clearer presentation of the terms introduced here. The coefficients in the
sum, by identifying terms with Eq.\ref{HamiltonianIdentify} correspond to
{
\begin{eqnarray}
	\begin{split}
	t_1 &\equiv V_{pp\pi} + \frac{1}{\epsilon_p - \epsilon_d}\left[2V_{pd_{z^2}}^2 + \frac{V_{pd_{xz}}^2}{3\zeta^2} - \frac{V_{pd_{xy}}^2}{36\zeta^4}\right],\\
	t_2 &\equiv \frac{1}{\epsilon_p - \epsilon_d}\left[V_{pd_{z^2}}^2 - \frac{V_{pd_{xz}}^2}{6\zeta^2} - \frac{V_{pd_{xy}}^2}{72\zeta^4}\right],\\
	t_3 &\equiv \frac{1}{\epsilon_p - \epsilon_d}\left[V_{pd_{z^2}}^2 - \frac{V_{pd_{xz}}^2}{3\zeta^2} + \frac{V_{pd_{xy}}^2}{36\zeta^4}\right],
	\end{split}
\end{eqnarray}}
where we have used that {$\vec{u}_{\, k} \cdot \vec{u}_{\, l} = \frac{1}{4}\left[(r_k \cdot \hat{R}_l + n_z^2)^2 - (\hat{r}_k \times \hat{R}_l)_z^2\right]$}.
This term represents a path that goes from the source to the gold and back and causes a shift in the energy $\epsilon_p$  of
{
\begin{equation}
	\frac{1}{\epsilon_p - \epsilon_d}\left(3V_{pd_{z^2}}^2 + \frac{V_{pd_{xz}}^2}{\zeta^2} + \frac{V_{pd_{xy}}^2}{12\zeta^4}\right)
\end{equation}}
that is added to $\tilde{\epsilon}$.

\subsubsection{Spin-Orbit Term $H_{SO}$}

Picking up the contributions to the SO terms, we can write it as
\begin{eqnarray}
	H_{SO} 	&=&  i\Delta_{xz, yz}^{SO}\sum_{l=1}^{6}(-1)^{l+1}s_za_{z,l} \nonumber\\
			& & + i\Delta_{xz,z^2}^{SO}\left[(-s_x + \sqrt{3}s_y)(a_{z,1} + a_{z,4}) \right. \nonumber\\
			& & \qquad\qquad - (s_x + \sqrt{3}s_y)(a_{z,2} + a_{z,5}) \nonumber\\
			& & \left. \qquad\qquad + 2s_x(a_{z,3} + a_{z,6})\right]\nonumber\\
			& & + i\Delta_{xz,xy}^{SO}\left[-2s_x(2b_{z,1} + b'_{z,1}) \right.\nonumber\\
			& & \qquad\qquad + (s_x + \sqrt{3}s_y)(2b_{z,2} + b'_{z,2}) \nonumber\\
			& & \left. \qquad\qquad + (s_x - \sqrt{3}s_y)(2b_{z,3} + b'_{z,3})\right]. \label{eq:so}
\end{eqnarray}
The first term on the right involves second neighbors couplings and SO interactions between $d_{xz}$ and $d_{yz}$ orbitals or between $d_{xy}$ and $d_{x^2-y^2}$ orbitals. The second term involves second neighbors and SO interactions between $d_{z^2}$ and $d_{xz}||d_{yz}$ orbitals. The last term involves first and third nearest neighbors and the SO interaction between $d_{xz}||d_{yz}$ and $d_{xy}||d_{x^2-y^2}$ orbitals. Note that the nearest neighbor interaction is twice as large as the one that involves third nearest neighbors. In Eq. (\ref{eq:so}) we find that the SO coefficients are
\begin{eqnarray}
	\begin{split}
	\Delta_{xz, yz}^{SO} &\equiv \frac{\xi_d}{2\sqrt{3}\zeta^2(\epsilon_p - \epsilon_d)^2}\left(V_{pd_{xz}}^2 - \frac{V_{pd_{xy}}^2}{6\zeta^2}\right),\\
	\Delta_{xz,z^2}^{SO} &\equiv \frac{\xi_dV_{pd_{xz}}V_{pd_{z^2}}}{2\zeta(\epsilon_p - \epsilon_d)^2},\\
	\Delta_{xz,xy}^{SO} &\equiv \frac{\xi_dV_{pd_{xz}}V_{pd_{xy}}}{6\sqrt{3}\zeta^3(\epsilon_p - \epsilon_d)^2},
	\end{split}
	\label{spinorbitdeltas}
\end{eqnarray}

were we have used the two following equalities
{
\begin{equation}
	\left(\vec{u}_{\, k} \times \vec{u}_{\, l}\right)_z = \frac{1}{2}(\hat{r}_k \times \hat{R}_l)_z(r_k \cdot \hat{R}_l + n_z^2),
\end{equation}}
and
{
{\footnotesize
\begin{eqnarray}
	\begin{split}
	& s_x\left(\hat{r}_k\times\vec{u}_j + \hat{r}_j\times\vec{u}_k\right)_z - s_y\left(\hat{r}_k\cdot\vec{u}_j + \hat{r}_j\cdot\vec{u}_k\right)\\
	& = \frac{1}{2}[(\hat{r}_k\cdot\hat{r}_j + n_z^2)(\vec{s}\times (\hat{r}_k + \hat{r}_j))_z \\ 
	& - (\hat{r}_k\times\hat{r}_j)_z(\vec{s}\cdot(\hat{r}_k - \hat{r}_j) - 2s_zn_z)].
	\end{split}
\end{eqnarray}
}}
The SO interaction does not cause a shift to the spectrum.

\section{Bloch Hamiltonian and spectral properties}
Having obtained the Hamiltonian in real space we derive the Bloch Hamiltonian in pseudo-spin or sublattice space. 

\subsection{Diagonal Term $H_{\rm AA}$}

First, we treat diagonal terms, i.e., terms that connect a site in the A sub-lattice with neighbors in the A sub-lattice as well, these are hops to second neighbors. We only show calculations for $H_{\rm AA}$ but the calculation is analogous to $H_{\rm BB}$ and leads to the same results. The full diagonal term {in the basis $\{A_{p_z}(1),A_{p_z}(2)\} \otimes \{\uparrow,\downarrow\}$ (where ($1$) is referred to the source atom in the sublattice A and ($2$) to the second neighbors of this atom also in the sublattice A)}, is written as:
\begin{widetext}
\begin{eqnarray}
	H_{\rm AA} &=& \left(\begin{array}{c c c c}
			-3t_2 - \Delta_{xz, yz}^{SO} & 0                        & 0                        & 0                        \\
			           0             & -3t_2 + \Delta_{xz, yz}^{SO} & 0                        & 0                        \\
			           0             & 0                        & -3t_2 + \Delta_{xz, yz}^{SO} & 0                        \\
			           0             & 0                        & 0                        & -3t_2 - \Delta_{xz, yz}^{SO}
			\end{array}\right) \nonumber\\
		&=& \varepsilon\sigma_0\otimes s_0 - \Delta_{xz, yz}^{SO}\sigma_z\otimes s_z,
\end{eqnarray}
\end{widetext}
where $\vec{\sigma}$ represents the pseudo-spin subspace and $\vec{s}$ the real spin subspace, $\varepsilon=-3t_2$ the chemical potential and $\sigma_0\otimes s_0 = \mathbb{I}_{4\times4}$ is the identity matrix (see Appendix B for a detailed calculation of these terms).

\subsection{Non-diagonal Term {$H^{\rm nd}$}}

Now, we treat non-diagonal terms, i.e., terms connecting A and B sub-lattices. These are hops to first and third neighbors. The full non-diagonal term {in the basis $\{A_{p_z},B_{p_z}\} \otimes \{\uparrow,\downarrow\}$}, is written as:
\begin{widetext}
\begin{eqnarray}
	H^{\rm nd} &=& \left(\begin{array}{c c c c}
		              0                & 0                             & \tilde{t}(p_x - ip_y)          & i\Delta_{xz,xy}^{SO}(\xi - 1) \\
		              0                & 0                             & -i\Delta_{xz,xy}^{SO}(\xi + 1) & \tilde{t}(p_x - ip_y)         \\
		    \tilde{t}(p_x + ip_y)      & i\Delta_{xz,xy}^{SO}(\xi + 1) & 0                              & 0                             \\
		-i\Delta_{xz,xy}^{SO}(\xi - 1) & \tilde{t}(p_x + ip_y)         & 0                              & 0
			\end{array}\right) \nonumber\\
	&=& (v_F + \tilde{v})\vec{\sigma}\cdot\vec{p} - \Delta_{xz,xy}^{SO}\left(\vec{s}\times\vec{\sigma}\right)_z,
\end{eqnarray}
\end{widetext}
where $\vec{\sigma}$ and $\vec{s}$ represent the spin and pseudo-spin subspaces, respectively. (see Appendix C for a detailed calculation of these terms).

The intrinsic SO coupling between $d_{xz}||d_{yz}$ and $d_{xy}||d_{x^2-y^2}$ orbitals conduces to a like-Rashba SO coupling in pristine graphene. Moreover, this kind of coupling between spin and pseudo-spin, $\left(\vec{s}\times\vec{\sigma}\right)_z$, produces flipping in the real spin when A-B hops take place.

\subsection{Low Energy Spectrum}

\begin{figure}[t]
	\centering
	\includegraphics[width=8.0cm,height=9.0cm]{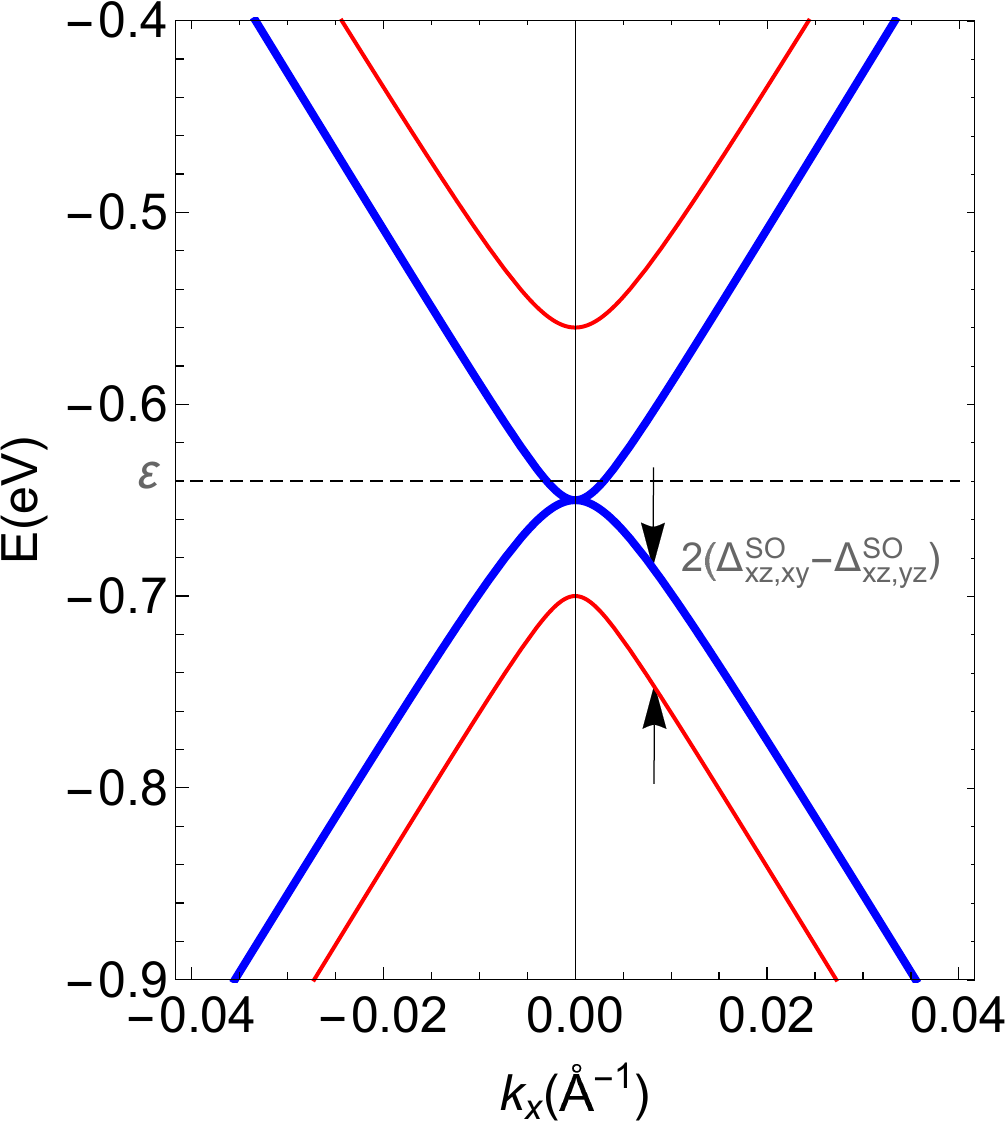}
	\caption{Low energy dispersion for the HCP graphene-gold system. The spectrum shows a gap between bands of $60$ meV due to the 
	$d_{xz(yz)}-d_{xy(x^2-y^2).}$} 
\label{spectrum}
\end{figure}

Diagonalizing the low energy Hamiltonian 
\begin{equation}
	H = \varepsilon\sigma_0\otimes s_0 + (v_F + \tilde{v})\hbar\vec{\sigma}\cdot\vec{k} - \Delta_{xz, yz}^{SO}\sigma_z\otimes s_z - \Delta_{xz,xy}^{SO}\left(\vec{s}\times\vec{\sigma}\right)_z
\end{equation}
we get the eigenvalues as
\begin{equation}
	\begin{split}
	\epsilon_1 &= \varepsilon -\Delta_{xz,xy}^{SO} - \sqrt{(v_F + \tilde{v})^2\hbar^2k^2  + (\Delta_{xz,xy}^{SO} - \Delta_{xz, yz}^{SO})^2}, \\
	\epsilon_2 &= \varepsilon -\Delta_{xz,xy}^{SO} + \sqrt{(v_F + \tilde{v})^2\hbar^2k^2 + (\Delta_{xz,xy}^{SO} - \Delta_{xz, yz}^{SO})^2},\\
	\epsilon_3 &= \varepsilon + \Delta_{xz,xy}^{SO} - \sqrt{(v_F + \tilde{v})^2\hbar^2k^2 + (\Delta_{xz,xy}^{SO} + \Delta_{xz, yz}^{SO})^2},\\
	\epsilon_4 &= \varepsilon + \Delta_{xz,xy}^{SO} + \sqrt{(v_F + \tilde{v})^2\hbar^2k^2 + (\Delta_{xz,xy}^{SO} + \Delta_{xz, yz}^{SO})^2},
	\end{split}
\end{equation}
where we have chosen the $\xi=+1$ value. Using the following magnitudes for the parameters, 
\begin{equation}
	\begin{split}
	\tilde{v} &= 0.27\times 10^6 ~{\rm m/s}, \\
	\varepsilon &= -0.64 ~{\rm eV}, \\
	\Delta_{xz, yz}^{SO} &= 10~{\rm meV},\\
	\Delta_{xz,xy}^{SO} &= 35~{\rm meV}.
	\end{split}
\end{equation}
{ These parameters depend on the site energies and the spin-orbit parameter of the gold d orbitals and the graphene-gold overlaps, (see Table~\ref{HCPparameters}). We note that in this configuration, we have a graphene-gold distance of $2.5\rm{\AA}$ for which we did not find reported values for the overlaps. Finally, we take the site energies of the $p_z$ orbitals of graphene as our reference energy, so $\epsilon_p = 0$, and $k_y=0$. With these values we get the spectrum shown in Figure \ref{spectrum}}. In the next section we will support the values of $\Delta_{xz, yz}^{SO}$ and $\Delta_{xz,xy}^{SO}$ on the basis of the elemental overlaps derived from
the model based on Eq.\ref{spinorbitdeltas}.

\begin{table}[h!]
\begin{center}
	\caption{Parameters used to obtain the low energy dispersion for the HCP graphene-gold system} \label{HCPparameters}
	\begin{tabular}{ l  c  c }
		\toprule
		Parameter         	& This work      		& Reference    								\\ \hline
		$\xi_d$      		& 0.3 eV 				& 0.65 eV \cite{Barreteau}  				\\
		$\epsilon_d$     	& 3.25 eV				& 3.27 eV 	\cite{Papaconstantopoulos}		\\
		$V_{pd\sigma}$ 	& -0.8 eV 				& --- 									\\
		$V_{pd\pi}$      		& 3.45 eV 			& --- 									\\ \toprule
	\end{tabular}
	\end{center}
\end{table}

{Figure~\ref{spectrum}} shows that the Dirac cone turns to a parabolic dispersion for both spin helicities and a spin-orbit splitting appears. The $\tilde{v}$ parameter changes the weight of the kinetic energy in the Hamiltonian. The denominator $\epsilon_p - \epsilon_d$ is negative, therefore, the correction to the kinetic energy {\bf increases} the velocity of the electrons. The chemical potential, $\varepsilon$, is negative so that graphene is doped by electrons from the gold layer. One can see that the SO gap, close to the Dirac point is identified by  the difference $2(\Delta_{xz,xy}-\Delta_{xz,yz})\sim 60$ meV. {This} figure recovers the results
of ab-initio studies in {references \cite{Marchenko, Krivenkov}} and adds detail of how this spectrum comes about from the hibridyzations of the $p_z$ graphene and
$d$ orbitals.


{
\section{ATOP configuration}}
{
In this section we analyze the Hamiltonian terms for the configuration in which each gold atom lies above each atom of one sublattice of graphene (the sublattice A in our case). This arrangement of the atoms is known as AC stacking (or ATOP configuration). }

{
In order to obtain the ATOP Hamiltonian, the procedure is very similar to the one illustrated in section~\ref{analytical}, and in the appendices~\ref{coeff},~\ref{diagelem}~and~\ref{ndiagelem}, where we took as our basis the $p_z$ orbitals of graphene and the $5d$ orbitals of gold. The difference in this case is that, due to the arrangement of the gold atoms, we can neglect the second and third neighbor hops. Then, we consider the Hamiltonian terms $H_{\rm AA(BB)}$ i.e. hops from one atom in the orbital $p_z$ in the sublattice A(B), to any of the $5d$ orbitals of gold, and then back to the orbital $p_z$ of the same atom in the sublattice A(B), renormalizing the site energy of graphene electrons in the orbital $p_z$, $\epsilon_{2p}$. For the Hamiltonian terms $H_{\rm AB}$, we consider hops $Ap_z-Bp_z$, either, directly through the coupling $V_{pp\pi}$ between first neighbors in the graphene's lattice, or going first to any of the $5d$ orbitals of gold and then to one of the first three neighbors $Bp_z$. With these ingredients, we obtained the expression for the low energy Hamiltonian}
{
\begin{equation}\label{atop1}
	H = \varepsilon\sigma_0\otimes s_0 + (v_F + \tilde{v})\hbar\vec{\sigma}\cdot\vec{k} - \Delta_{xz,z^2}^{SO}\left(\vec{s}\times\vec{\sigma}\right)_z+h_{z0} \sigma_z\otimes s_0,
\end{equation}
where
\begin{equation}\label{atop2}
	\begin{split}
	\tilde{v} &= \frac{\sqrt{3}a}{2\hbar}\frac{V'_{pd_{z^2}}V_{pd_{z^2}}}{(\epsilon_p-\epsilon_d)}, \\
	\varepsilon &= \frac{t_1+t_2}{2}, \\
	\Delta_{xz,z^2}^{SO} &= \frac{3\xi_dV'_{pd_{xz}}V_{pd_{z^2}}}{2\zeta(\epsilon_p - \epsilon_d)^2},\\
	h_{z0} &= \frac{t_1-t_2}{2},
	\end{split}
\end{equation}}
{with}
{\begin{equation}\label{atop3}
	\begin{split}
	t_1 &= -\frac{V_{pd_{z^2}}^2}{(\epsilon_p-\epsilon_d)}, \\
	t_2 &=\frac{1}{(\epsilon_p-\epsilon_d)}\Big[ -V_{pd_{z^2}}^{'2}-\frac{V_{pd_{xy}}^{'2}}{3\zeta^4}-\frac{V_{pd_{xz}}^{'2}}{\zeta^2}\Big].
	\end{split}
\end{equation}}

{In Eqs.~\ref{atop2}~and~\ref{atop3}, the primed V's are used to distinguish the overlaps between the $Bp_z$ orbitals with the $5d$ orbitals of gold, while the unprimed V's are used for the overlaps between the $Ap_z$ orbitals with the $5d$ orbitals of gold. This distinction is necessary in the configuration ATOP, where the atoms of the sublattice A, by symmetry, have nonzero overlap only with the orbital $5d_{z^2}$ of gold, while the atoms of the sublattice B have nonzero overlap with all the $5d$ orbitals of gold. Then the cosine directors graphene-gold are different for both sublattices.}

\begin{figure}[t]
	\centering
	\includegraphics[width=8.0cm,height=7.5cm]{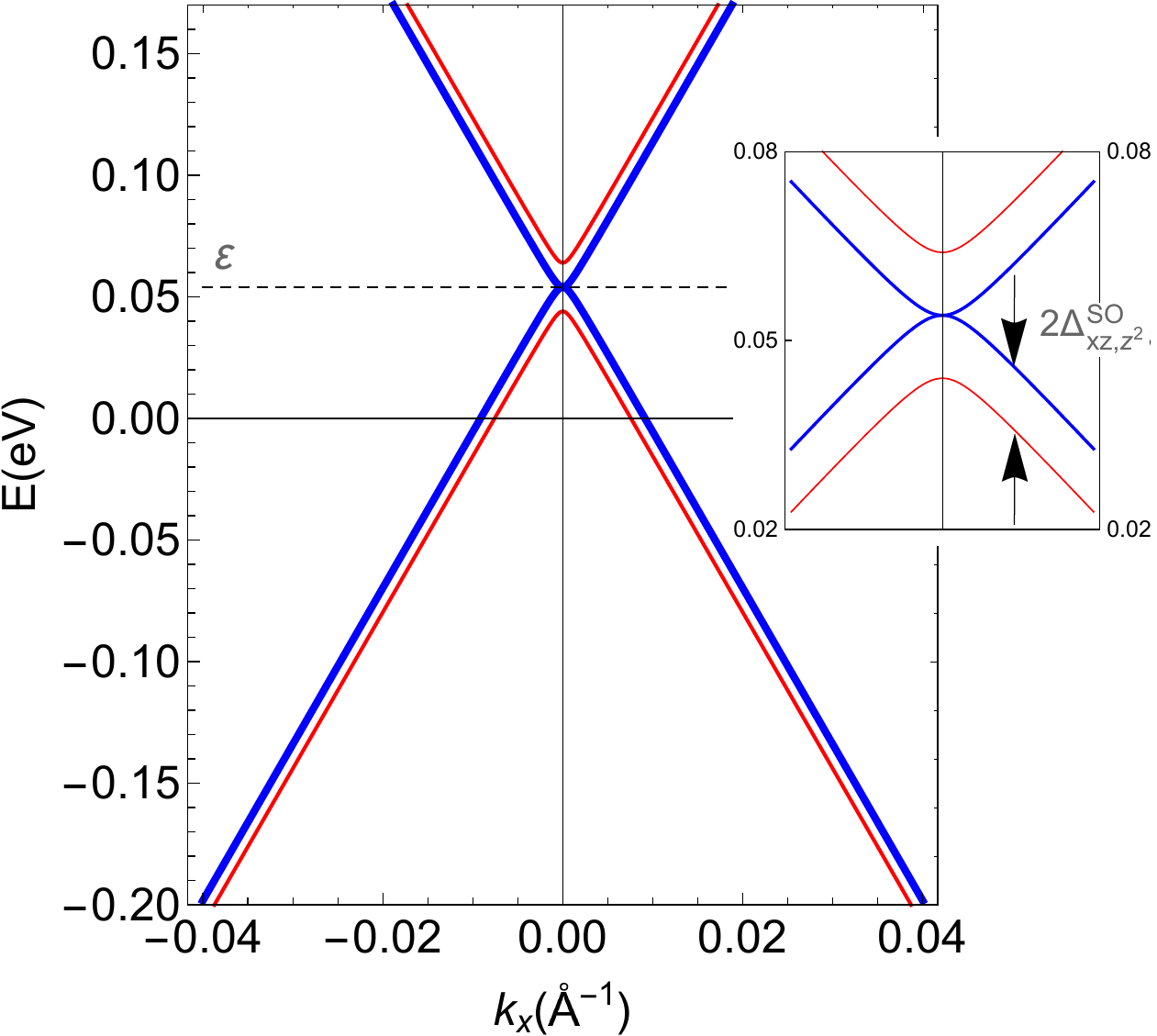}
	\caption{Low energy dispersion for the ATOP graphene-gold system. The spectrum shows a gap between bands of $10$ meV due to the 
	$d_{xz(yz)}-d_{z^2}$. We took the values: $\tilde{v}=-3.85\times 10^3$m/s, $\varepsilon=0.054$eV, $\Delta_{xz,z^2}^{SO}=5$meV and $h_{z0}=0$. These values were obtained taking  the site energy and the spin orbit parameters of the gold's d orbitals, and the graphene-gold overlaps reported in table~\ref{ATOPparameters}, and $\epsilon_p = 0$ as our reference. The inset shows that the SO gap, close to the Dirac point is $2\Delta_{xz,z^2}$.} 
\label{ATOP}
\end{figure}

{Going back to Eq.~\ref{atop1}, we see that we have, as before, a first term corresponding to the chemical potential, and the second is the kinetic term. The third is a Rashba like term originated from the intrinsic SO coupling between $d_{xz}||d_{yz}$ and $d_{z^2}$ orbitals. Finally the fourth term comes from the symmetry breaking between the A and B sublattices, as a result of the positioning of the gold atoms in the ATOP configuration. As discussed in ref. \cite{Marchenko}, this symmetry breaking can be neglected at gold-graphene distances larger than $2.5\rm{\AA}$, { and given that the graphene-gold equilibrium distance reported for this structure is $3.3$\AA \cite{Marchenko}}, we can neglect the fourth term of the Equation~\ref{atop1} and diagonalize it, obtaining the bands shown in Figure~\ref{ATOP}.}
\begin{table}
\begin{center}
	\caption{Parameters used to obtain the low energy dispersion for the ATOP graphene-gold system} \label{ATOPparameters}
	\begin{tabular}{ l  c  c }
		\toprule
		Parameter         	& This work      		& Reference    							\\ \hline
		$\xi_d$      		& 0.3 eV 				& 0.65 eV \cite{Barreteau}   			\\
		$\epsilon_d$     	& 3.25 eV				& 3.27 eV \cite{Papaconstantopoulos}	\\
		$V_{pd\sigma}$ 	& -0.28 eV 			& -0.24 eV \cite{Qin}				\\
		$V_{pd\pi}$      		& 0.23 eV 			& -0.16 eV \cite{Qin}				\\ \toprule
	\end{tabular}
	\end{center}
\end{table}

{Figure~\ref{ATOP} shows that, as in the HCP case, the Dirac cone turns to a parabolic dispersion for both spin helicities and a spin-orbit splitting appears. Also, the $\tilde{v}$ parameter changes the weight of the kinetic energy in the Hamiltonian, with $\tilde{v}<0$, therefore, the correction to the kinetic energy {\bf decreases} the velocity of the electrons. The chemical potential, $\varepsilon$, is {\bf positive} so that graphene is doped by {\bf holes} from the gold layer. In this case the SO gap, close to the Dirac point is $2\Delta_{xz,z^2}^{SO}\sim 10$ meV, a much smaller effect than in the AC configuration, leading to the conclusion that this is not a convenient register
to enhance the SO coupling. This figure recovers the results of ab-initio studies in reference \cite{Marchenko} for the ATOP configuration. }

\section{Supercell treatment}
As discussed by Marchenko et. al. \cite{Marchenko}, the register of one gold atom at each graphene plaquette at a distance of $2.5\rm{\AA}$ generates a giant SO coupling, but only at a cost of $\sim 1$ eV in repulsion energy relative to the equilibrium separation. So this geometry is considered unrealistic for the experimental situation. { However, Krivenkov et. al. \cite{Krivenkov}, found that the intercalation of Au on Graphene/Ni(111) occurs in the form of nanoclusters with different periodicities and sizes, which coexist with a continuous monolayer of gold. So, the giant SOC is attributed to the reduction of the equilibrium graphene-gold distance (to $2.35\rm{\AA}$), due to the attraction of graphene to the Ni exposed between the Au clusters. Here we discuss the analytical treatment of these clusters of gold above graphene. To preserve the simplicity of the model, we only consider clusters of one atom of Au in the HCP register, which is the configuration found by STM and DFT in reference \cite{Krivenkov}. This setup has essentially the same SO enhancement, according to ab initio studies \cite{Marchenko, Krivenkov}}. 

\begin{figure}[h]
	\centering
	\includegraphics[width=5.7cm,height=11.0cm]{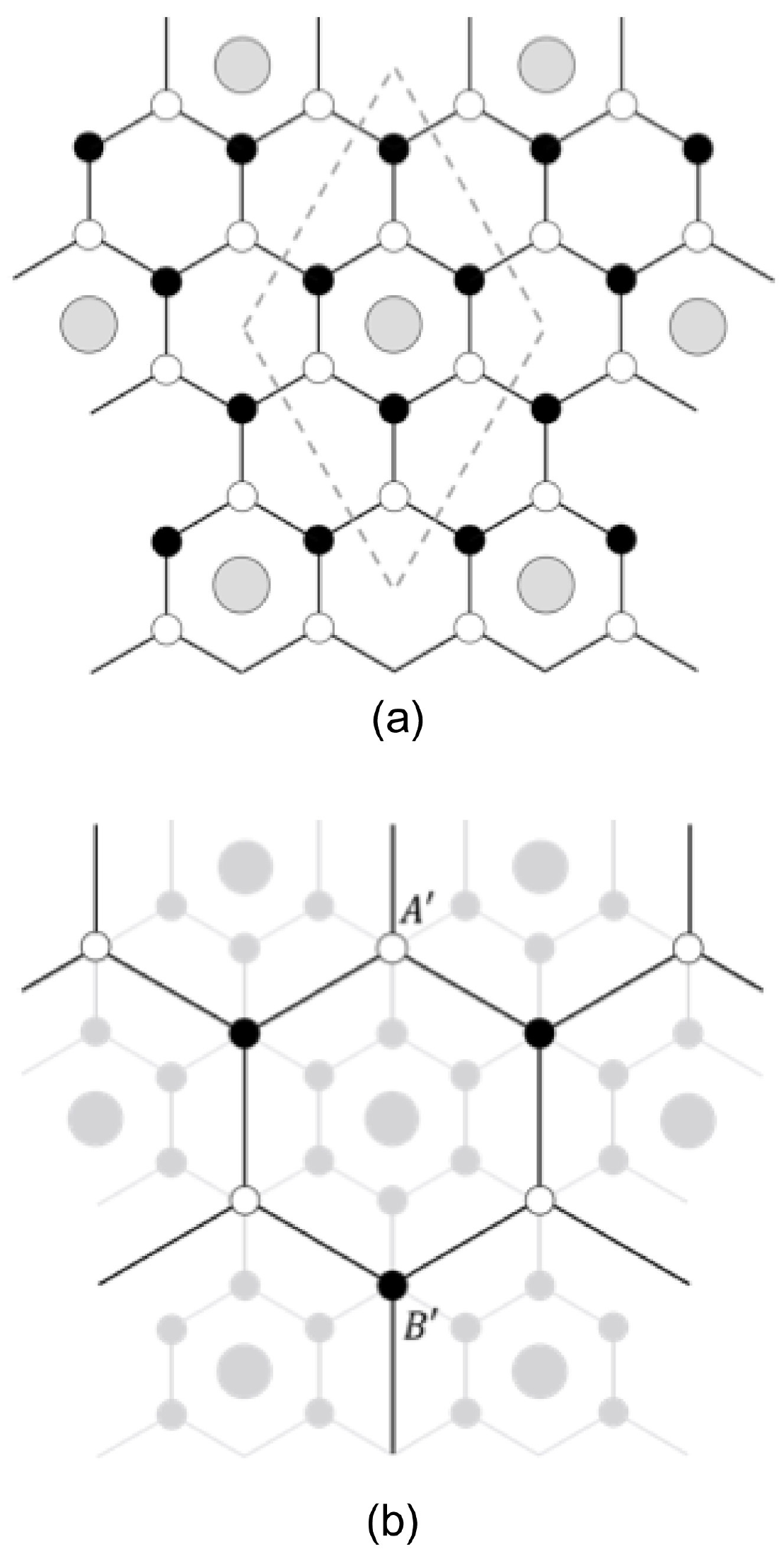}
	\caption{a) Graphene on a diluted gold surface. The dashed diamond delimits the unitary supercell formed by a gold atom and eight graphene Carbons, and b) the equivalent honeycomb lattice, where the black lines are the effective links between sites. Note that the outer hexagon (black) maps onto
the inner (grey) hexagon through $p_z-p_z$ overlaps. The red (online) arrows depict one of the processes involved in Eq.\ref{DecimatedCoupling}.
\label{diluted}}
\end{figure}

We discuss here the spin-orbit enhancement when gold atoms are diluted in such a way that each one is surrounded by six empty graphene's plaquettes (see Figure \ref{diluted} panel a). The corresponding tight-binding model for this situation can be readily obtained by defining a graphene-gold supercell approach. The unit supercell is formed by eight graphene atoms and a gold atom (atoms inside the dashed rhombus in Figure \ref{diluted} a). This supercell can be seen as an equivalent honeycomb lattice with $\rm A'$, $\rm B'$ sublattices. Note that the internal hexagon of the supercell maps onto the external hexagon by overlaps between graphene $p_z$ orbitals in the outer hexagon and the hybridized $p_z-d$ orbital in the inner ring. 

If we reduce the coupled equations using lowest order perturbations theory, as previously, to obtain the coupling between the outer and the inner hexagons we obtain the following expression
\begin{widetext}
\begin{equation}
\left[\epsilon -\epsilon_p-3\frac{V^2_{pp\pi}}{\epsilon_p-\tilde\epsilon_{pd}}\right ] A_{z0}=\sum_{l=1}^{3}\sum_{j=1}^3\frac{V^2_{pp\pi}}{(\epsilon_p-\epsilon_{pd})(\epsilon_p-\tilde\epsilon_{pd})}\left [-(\vec{n}_{l}\cdot\vec{n}_{lj})\frac{\tilde{V}^2_{pp\pi}}{\epsilon_p-\tilde\epsilon_d}+(\vec{u}_{l}\cdot\vec{u}_{lj})\frac{{V}^2_{pdxy}}{\epsilon_p-\tilde\epsilon_d}\right ]B_{zlj},
\label{DecimatedCoupling}
\end{equation}
\end{widetext}
where we distinguish outer ring sites by uppercase letters $A_{z0}$, $B_{zlj}$. The first subindex denotes the $p_z$ orbital on the graphene A sublattice sites, and B sublattice site respectively.   $\epsilon_{pd}$ denotes a graphene site energy close to the gold atom, while $\tilde\epsilon_{pd}$ involves additional corrections from the gold atom orbital overlaps. Fig.\ref{diluted} panel b) depicts the process
represented by  Eq.\ref{DecimatedCoupling}. The quantity in the bracket on the left is the expression for the A-B coupling when all graphene plaquettes are filled by gold. So the prefactor is the additional overlaps involved for the transfer between the outer hexagon and the inner hexagon and it renormalizes all couplings
of the larger hexagon which now has gold in every plaquette. As can be seen, this is just one step in a renormalization group process in real space. 

We then define
\begin{equation}
\beta=\frac{V^2_{pp\pi}}{(\epsilon_p-\epsilon_{pd})(\epsilon_p-\tilde\epsilon_{pd})}.
\end{equation}
We have proven this result in detail for all the couplings in an extended derivation which is not included here. Nevertheless, it can be derived by a more direct
decimation argument\cite{PastawskiMedina}. As all $p_z-p_z$ couplings are obtained on the basis of the shortest Feynman paths,
the dimensionless $\beta$ factor will renormalize the kinetic energies and the SO couplings.

\begin{figure}[t]
	\centering
	\includegraphics[width=8.0cm,height=9.0cm]{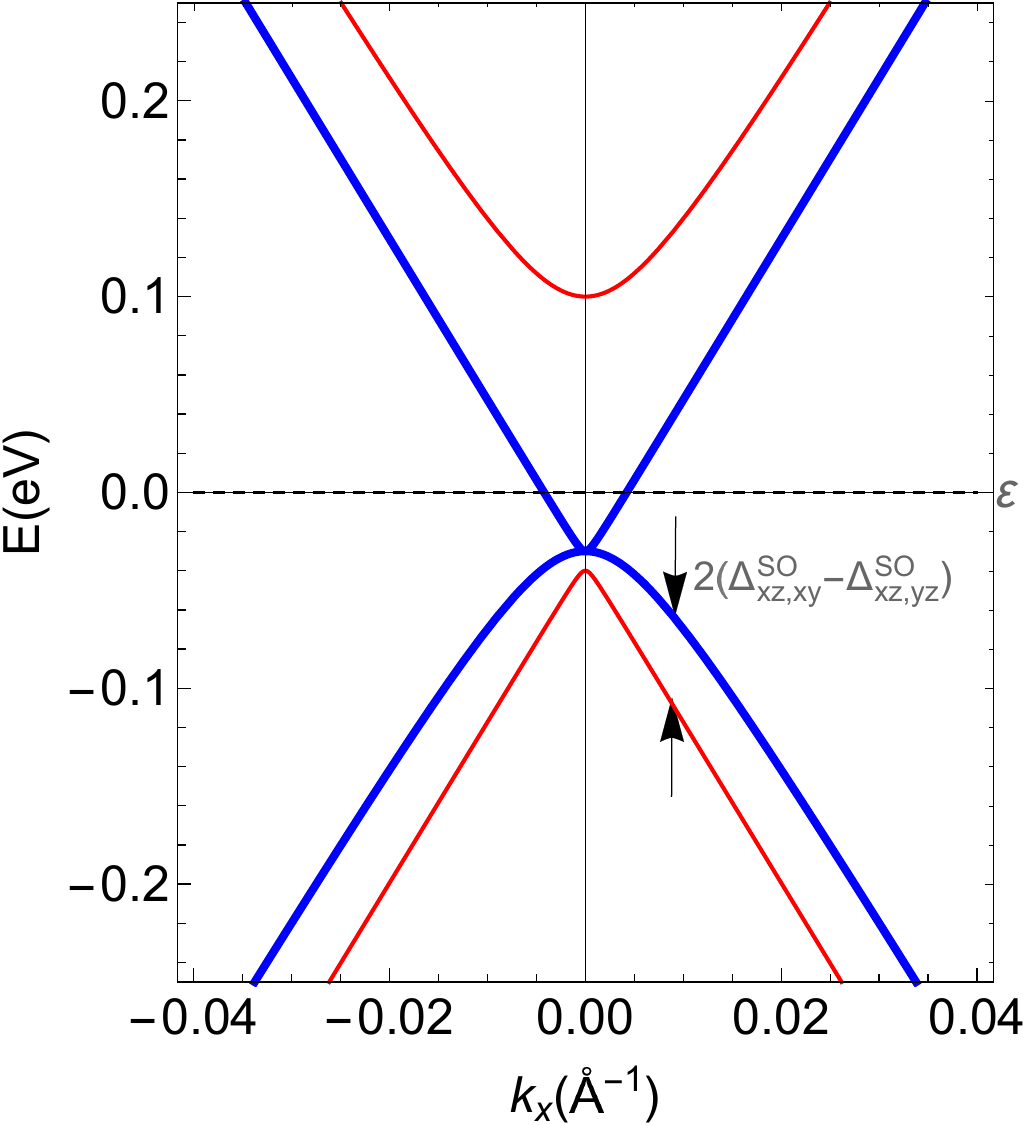}
	\caption{Low energy dispersion for the HCP system of diluted gold on graphene. The spectrum shows a gap between inferior bands of $50$ meV due to the $d_{xz(yz)}$-$d_{xy(x^2-y^2)}$ spin-orbit interaction. Since the  magnitude of $\Delta_{xz, yz}^{SO}$ is comparable to the magnitude of $\Delta_{xz,xy}^{SO}$, the Dirac cone is broken at the K point and a gap appears. 
\label{spectrum2}}
\end{figure}
\begin{figure}[h]
	\centering
	\subfigure[]{\includegraphics[width=.52\textwidth]{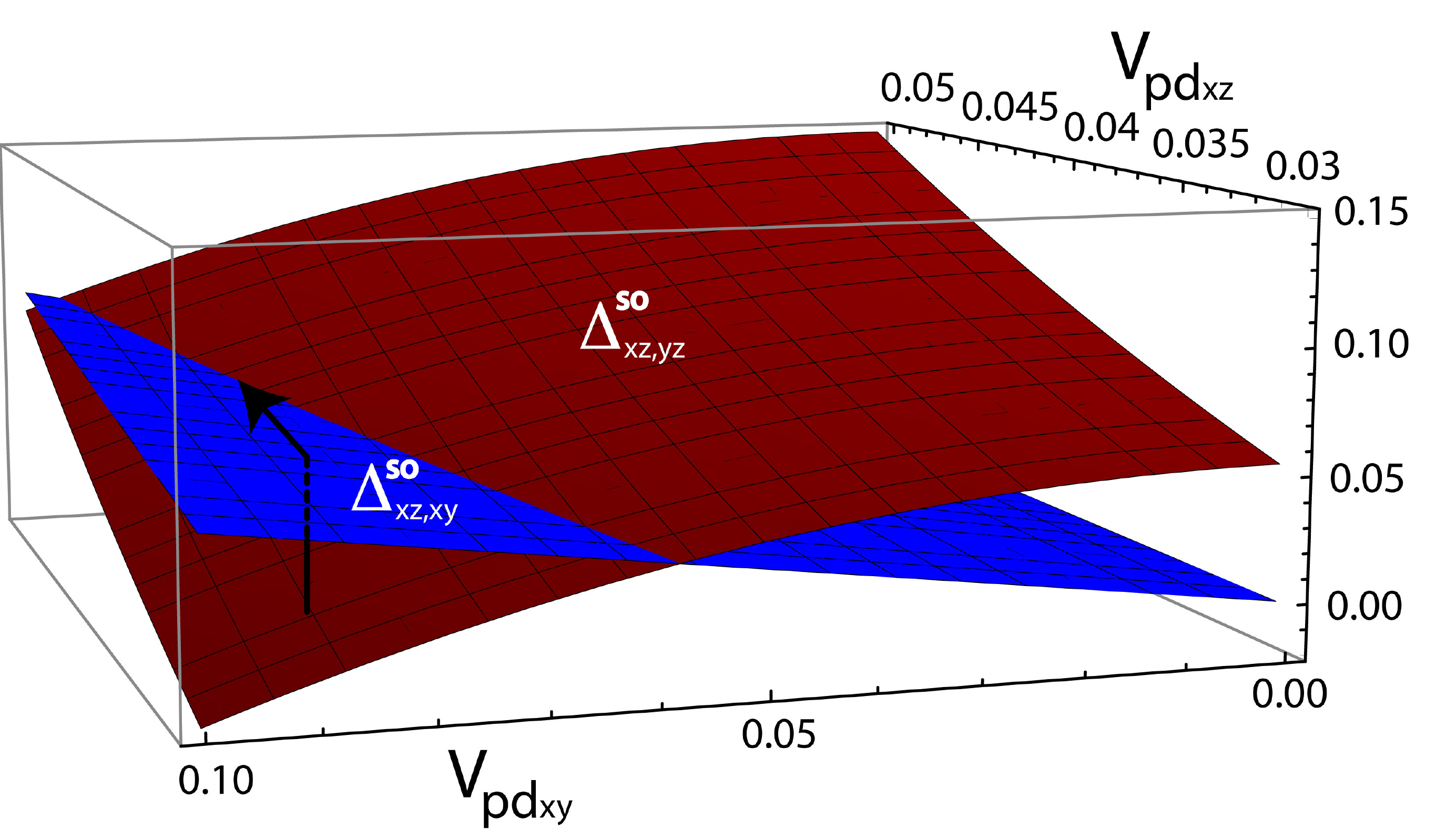}}
	\subfigure[]{\includegraphics[width=.5\textwidth]{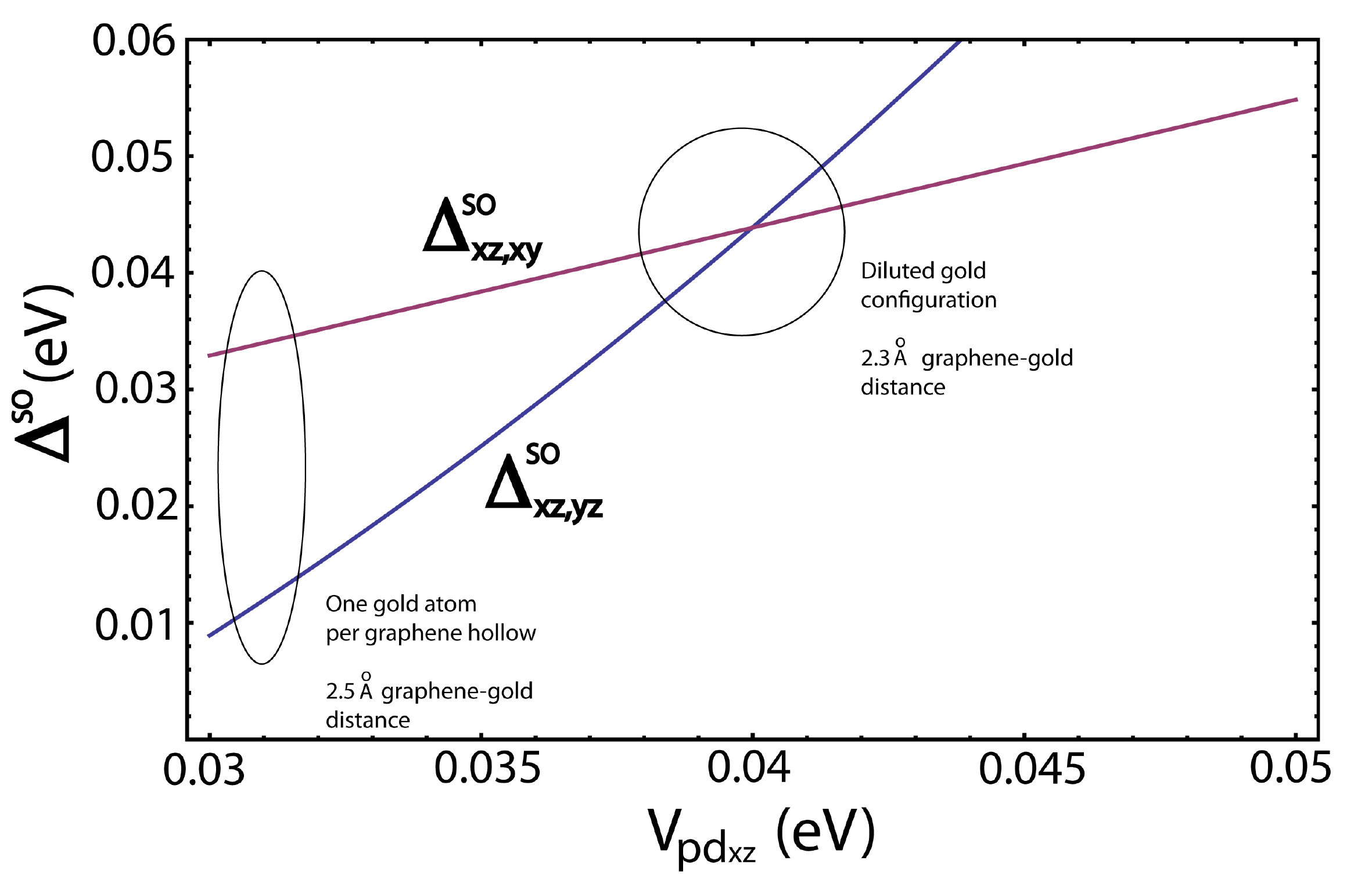}}
	\caption{a) Values of SO parameters as a function of the overlaps $V_{pd_{xz}}$ and $V_{pd_{xz}}$ in a physical parameter range. The arrow indicate
	a surmised course of evolution of the SO parameters when the graphene is moved closer to the gold surface. b) A cut of the surface in part (a) for $V_{pd_{xz}}$ fixed. For the larger separations ($2.5$\AA) the parameters correspond to the values that fit the indiluted gold surface, while for the smaller separation
	($2.3$\AA) the parameters approach those of the diluted gold surface. The circle and ellipse indicate the error bars of the model.
\label{OverlapInterference}}
\end{figure}

If the new spectrum for the diluted gold surface only depended on a rescaling by $\beta$, then there would be no qualitative changes
to Fig.\ref{spectrum}, while results from ab-initio calculations show a shift in the chemical potential of ~0.5 eV  and the SO gap for the valence band, almost vanishes
at the Dirac point. On the other hand, the SO gap is basically preserved when $k_x$ vector is large enough. The spectrum is a combination of 
a preserved Dirac cones for the valence band with a selected helicity and the opposite helicity has a quadratic dispersion\cite{Marchenko}.

Our results readily reproduce the expected spectrum if, besides the scaling of the couplings by $\beta$ we also take into account changes in the
gold-graphene distance {(in agreement with the discussion presented in reference \cite{Krivenkov})}. The following parameters can be chosen to fit the ab-initio spectrum
\begin{equation}
	\begin{split}
	\tilde{v} &= 0.27\times 10^6 ~{\rm m/s}, \\
	\varepsilon &\sim 0. ~{\rm eV}, \\
	\Delta_{xz, yz}^{SO} &= 30~{\rm meV},\\
	\Delta_{xz,xy}^{SO} &= 35~{\rm meV}.
	\end{split}
\end{equation}
How the parameter $\Delta_{xz,xy}^{SO}$ is preserved while $\Delta_{xz, yz}^{SO}$ increases can be understood by { an}
interference effect between the overlaps of the graphene {\bf$p_{z}$} and gold $d$ orbitals. Figure \ref{OverlapInterference}(a)
shows the behavior of the two SO coupling parameters as a function of the local orbital overlaps of $p_z$ graphene and
$d_{xy}, d_{xz}$ orbitals according to Eq.\ref{spinorbitdeltas}. One can see that if we follow the parameter choices along
the arrow on the  sheet describing $\Delta_{xz,xy}^{SO}$ we begin with a parameter set corresponding to the indiluted gold covering,
and ends, within error bars (depicted by circle and ellipse), in the final diluted gold, covering and a shorter distance between
tha Gold and the graphene surface. { This last observation is also in agreement with the results shown in reference \cite{Krivenkov} discussed
previously.}

Fig.\ref{OverlapInterference}(b) shows the cut along the arrow in panel (a) that fixes the value of $\Delta_{xz,xy}^{SO}$ and
approaches the intersection with the sheet describing $\Delta_{xz,yz}^{SO}$, showing good agreement with the selected parameters
producing Fig.\ref{spectrum2}.

{Finally, the DFT model presented in reference \cite{Krivenkov}, also shows that the gold clusters produce a buckling on the graphene. This buckling generates a gap at the Dirac point and a non zero out of plane polarization in the vicinity of the gap, but it does not affect appreciably the spin orbit coupling. So, we do not include this effect in our supercell model.}


\section{Summary and conclusions}

We have discussed a simple analytical tight binding model for gold over graphene in three emblematic
registries; ATOP, undiluted gold covering in the HCP register and the diluted gold covering in the HCP register. We have used
lowest order perturbation theory and the Slater-Koster tight binding approach to arrive at an effective Hamiltonian for the
Graphene perturbed by proximity effects from the Gold surface. The main motivation of this work is to understand the mechanisms
by which the SOC in graphene can be enhanced by way of proximity effects that do not interfere with properties 
like electron mobility on the graphene sheet.

We find excellent agreement in deriving the spin dependent band structure with both experimental findings and 
detailed DFT studies in ref. {\cite{Marchenko, Krivenkov}. The model correctly describes both the SO coupling 
in all the registries above and the chemical potential that makes for a non zero electron doping on the graphene after 
adjusting a minimal number of parameters.  Our results reveal, in detail, the interplay between
graphene {\bf $p_{z}$} orbitals and gold {\bf $5d$} orbitals that give rise to the SO interaction in a non-trivial way; the
result of interfering contributions from different atomic SO couplings. The interplay between orbital overlap and graphene-gold 
distance renders results such as the preservation of the SO gap {in the presence of a diluted gold HCP covering}. 
Finally we have developed a renormalization group argument to deal with large primitive cells which would be very cumbersome
under directo Slater-Koster treatment. The results show that the diluted Gold registry is almost as effective as the
undiluted case as shown in the DFT calculation. The diluted case, as has been argued in the literature, is more
relevant to the experimental realization.


\acknowledgements{We thank Bertrand Berche for illuminating discussions. F.M. and M.P. acknowledges the support of PAPIIT-UNAM through the project IN111317. M.P. is also grateful to Conacyt-SENER for their support.}

\appendix
\section{Gold-Graphene Hopping Integrals and $c_{\mu,k}$ Coefficients}\label{coeff}
The gold-graphene hopping integrals for the $5d$ gold orbitals are:
{\footnotesize
\begin{eqnarray}\label{hintegrals1}
(\epsilon-\epsilon_d) c_{z^2,k}	&=&  -i\sqrt{3}s_y\xi_dc_{xz,k} + i\sqrt{3}s_x\xi_dc_{yz,k} \nonumber\\
								& & + \sum_{l}E_{z^2,z}^{l}g_{z,l}\nonumber\\
						&=&  -i\sqrt{3}s_y\xi_dc_{xz,k} + i\sqrt{3}s_x\xi_dc_{yz,k} \nonumber\\
								& & + V_{pd_{z^2}} \sum_{l}g_{z,l}. \label{cz2_eq}
\end{eqnarray}
\begin{eqnarray}\label{hintegrals2}
	(\epsilon-\epsilon_d) c_{xz,k} 	&=& i\sqrt{3}s_y\xi_dc_{z^2,k} - is_z\xi_dc_{yz,k} - is_y\xi_dc_{x^2-y^2,k} \nonumber\\
								   	& & + is_x\xi_dc_{xy,k} + \sum_{l}E_{xz,z}^{l}g_{z,l} \nonumber\\
								   	&=& i\sqrt{3}s_y\xi_dc_{z^2,k} - is_z\xi_dc_{yz,k} - is_y\xi_dc_{x^2-y^2,k} \nonumber\\ 
								   	& & + is_x\xi_dc_{xy,k} - V_{pd_{xz}} \sum_{l}n_{lx} g_{z,l},
\end{eqnarray}
\begin{eqnarray}\label{hintegrals3}
	(\epsilon-\epsilon_d) c_{yz,k} 	&=& -i\sqrt{3}s_x\xi_dc_{z^2,k} + is_z\xi_dc_{xz,k} - is_x\xi_dc_{x^2-y^2,k} \nonumber\\
									& & - is_y\xi_dc_{xy,k} + \sum_{l}E_{yz,z}^{l}g_{z,l} \nonumber\\
									&=& -i\sqrt{3}s_x\xi_dc_{z^2,k} + is_z\xi_dc_{xz,k} - is_x\xi_dc_{x^2-y^2,k} \nonumber\\ 
									& & - is_y\xi_dc_{xy,k} - V_{pd_{xz}}\sum_{l}n_{ly} g_{z,l},
\end{eqnarray}
\begin{eqnarray}\label{hintegrals4}
	(\epsilon-\epsilon_d) c_{x^2-y^2,k}	&=& is_y\xi_dc_{xz,k} + is_x\xi_dc_{yz,k} - i2s_z\xi_dc_{xy,k} \nonumber\\
										& & \sum_{l}E_{x^2-y^2,z}^{l}g_{z,l} \nonumber\\
										&=& is_y\xi_dc_{xz,k} + is_x\xi_dc_{yz,k} - i2s_z\xi_dc_{xy,k}\nonumber\\
										& & + V_{pd_{xy}}\sum_{l}u_{lx} b_{z,l},
\end{eqnarray}
\begin{eqnarray}\label{hintegrals5}
	(\epsilon-\epsilon_d) c_{xy,k}	&=& -is_x\xi_dc_{xz,k} + is_y\xi_dc_{yz,k} + i2s_z\xi_dc_{x^2-y^2,k}\nonumber\\
									& & + \sum_{l}E_{xy,z}^{l}g_{z,l} \nonumber\\
									&=& -is_x\xi_dc_{xz,k} + is_y\xi_dc_{yz,k} + i2s_z\xi_dc_{x^2-y^2,k}\nonumber\\
									& & + V_{pd_{xy}}\sum_{l}u_{ly} g_{z,l}. \label{cxy_eq}
\end{eqnarray}
}

In order to get the expansion coefficients $c_{\mu,k}$ in terms of the $g_{z,l}$, we solved the system of coupled equations keeping terms up to first order in $\xi_d$,
the SO coupling parameter, considered as a perturbation. The terms are written as
{\footnotesize
\begin{eqnarray}\label{ccoef1}
	c_{z^2,k}	&=& \frac{V_{pd_{z^2}}}{\epsilon-\epsilon_d}\sum_{l}g_{z,l} 
					+ \frac{i\sqrt{3}\xi_dV_{pd_{xz}}}{(\epsilon-\epsilon_d)^2}\sum_{l}\left(n_{lx} s_y - n_{ly} s_x\right)g_{z,l}\nonumber\\
\end{eqnarray}
\begin{eqnarray}\label{ccoef2}
	c_{xz,k}	&=& -\frac{V_{pd_{xz}}}{\epsilon-\epsilon_d}\sum_{l}n_{lx} g_{z,l} 
					+ \frac{i\xi_d}{(\epsilon-\epsilon_d)^2} \left[\sqrt{3}s_yV_{pd_{z^2}}\sum_{l}g_{z,l} \right. \nonumber\\
				& & \left. + s_z V_{pd_{xz}} \sum_{l}n_{ly} g_{z,l}
					+ V_{pd_{xy}} \sum_{l} \left(s_x u_{ly} - s_y u_{lx} \right)g_{z,l} \right] \nonumber\\
\end{eqnarray}
\begin{eqnarray}\label{ccoef3}
	c_{yz,k}	&=& - \frac{V_{pd_{xz}}}{\epsilon-\epsilon_d}\sum_{l}n_{ly} g_{z,l}
					- \frac{i\xi_d}{(\epsilon-\epsilon_d)^2}\left[\sqrt{3}s_x V_{pd_{z^2}}\sum_{l}g_{z,l} \right. \nonumber\\
				& & \left. + s_z V_{pd_{xz}} \sum_{l} n_{lx} g_{z,l}
					+ V_{pd_{xy}}\sum_{l} \left(s_x u_{lx} + s_y u_{ly} \right)g_{z,l} \right] \nonumber \\
\end{eqnarray}
\begin{eqnarray}\label{ccoef4}
	c_{x^2-y^2,k}	&=& \frac{V_{pd_{xy}}}{\epsilon-\epsilon_d}\sum_{l}u_{lx} g_{z,l} 
						- \frac{i\xi_d}{(\epsilon-\epsilon_d)^2}\left[ 2s_zV_{pd_{xy}}\sum_{l} u_{ly} g_{z,l} \right. \nonumber\\
					& &	\left. +  V_{pd_{xz}}\sum_{l}\left(n_{lx} s_y + n_{ly} s_x\right)g_{z,l} \right]
\end{eqnarray}
\begin{eqnarray}\label{ccoef5}
	c_{xy,k}	&=& \frac{V_{pd_{xy}}}{\epsilon-\epsilon_d}\sum_{l} u_{ly} g_{z,l} 
					+ \frac{i\xi_d}{(\epsilon-\epsilon_d)^2}\left[ 2s_zV_{pd_{xy}}\sum_{l} u_{lx} g_{z,l} \right. \nonumber\\
				& & \left. + V_{pd_{xz}}\sum_{l} \left(n_{lx} s_x - n_{ly} s_y\right) g_{z,l} \right].
\end{eqnarray}
}

\section{Diagonal Elements: Kinetic and SO Contributions of the Bloch Hamiltonian}\label{diagelem}
\subsection{Kinetic Contribution}

\begin{eqnarray}
	H_{AA}^{K} 	&=& t_2\left[e^{i\vec{k}\cdot(-\vec{\delta}_1 + \vec{\delta}_2)}
					+e^{i\vec{k}\cdot(-\vec{\delta}_1 + \vec{\delta}_3)}
					+e^{i\vec{k}\cdot(-\vec{\delta}_2 + \vec{\delta}_3)} \right.\nonumber\\
				& &	\left. \quad +e^{i\vec{k}\cdot(-\vec{\delta}_2 + \vec{\delta}_1)}
					+e^{i\vec{k}\cdot(-\vec{\delta}_3 + \vec{\delta}_1)}
					+e^{i\vec{k}\cdot(-\vec{\delta}_3 + \vec{\delta}_2)}\right]\nonumber\\
				&=& t_2g_1(\vec{k}) \label{h_aa_k},
\end{eqnarray}
where 
\begin{displaymath}
	\vec{\delta}_1 = \left(0,\frac{a}{\sqrt{3}}\right),
	\vec{\delta}_2 = \left(\frac{a}{2}, -\frac{a}{2\sqrt{3}}\right),
	\vec{\delta}_3 = \left(-\frac{a}{2}, -\frac{a}{2\sqrt{3}}\right).
\end{displaymath}
and
\begin{equation}
	g_1(\vec{k}) \equiv 2 \left[\cos(a k_x) + 2 \cos\left(\frac{a k_x}{2}\right) \cos\left(\frac{\sqrt{3} a k_y}{2}\right)\right].
\end{equation}
Expanding $g_1(\vec{k})$ around Dirac's point, $\vec{K}_\xi=\left(\xi\frac{4\pi}{3a}, 0\right)$, up to zero order in $\vec{p} = \hbar\vec{k} - \hbar\vec{K}_\xi$, to study low energy electrons, we get
\begin{equation}
	g_1(\vec{k}) \sim -3.
\end{equation}

Therefore,
\begin{equation}
	H_{AA}^T \sim -3t_2.
\end{equation}
\\
\subsection{Spin-Orbit Contribution}
For the diagonal SO term we have
\begin{eqnarray}
	H_{AA}^{SO} &=& i\Delta_{xz,z^2}^{SO}\left[(-s_x + \sqrt{3}s_y)(e^{i\vec{k}\cdot(-\vec{\delta}_1 + \vec{\delta}_2)} + e^{i\vec{k}\cdot(-\vec{\delta}_2 + \vec{\delta}_1)}) \right.\nonumber\\
				& & \qquad\qquad - (s_x + \sqrt{3}s_y)(e^{i\vec{k}\cdot(-\vec{\delta}_1 + \vec{\delta}_3)} + e^{i\vec{k}\cdot(-\vec{\delta}_3 + \vec{\delta}_1)}) \nonumber\\
				& & \qquad\qquad \left. + 2s_x(e^{i\vec{k}\cdot(-\vec{\delta}_2 + \vec{\delta}_3)} + e^{i\vec{k}\cdot(-\vec{\delta}_3 + \vec{\delta}_2)})\right] \nonumber\\
				& & + i\Delta_{xz, yz}^{SO}s_z\left[e^{i\vec{k}\cdot(-\vec{\delta}_1 + \vec{\delta}_2)}
					-e^{i\vec{k}\cdot(-\vec{\delta}_1 + \vec{\delta}_3)}
					+e^{i\vec{k}\cdot(-\vec{\delta}_2 + \vec{\delta}_3)} \right.\nonumber\\
				& & \left. \qquad\qquad -e^{i\vec{k}\cdot(-\vec{\delta}_2 + \vec{\delta}_1)}
					+e^{i\vec{k}\cdot(-\vec{\delta}_3 + \vec{\delta}_1)}
					-e^{i\vec{k}\cdot(-\vec{\delta}_3 + \vec{\delta}_2)}\right]\nonumber\\
	&=& i\Delta_{xz,z^2}^{SO}\left(s_xg_2(\vec{k}) + s_yg_3(\vec{k})\right) + \Delta_{xz, yz}^{SO}s_zg_4(\vec{k}),
\end{eqnarray}
where 
\begin{eqnarray}
	g_2(\vec{k}) &\equiv& 4\left[\cos(ak_x) - \cos\left(\frac{ak_x}{2}\right)\cos\left(\frac{\sqrt{3}ak_y}{2}\right)\right],\nonumber\\
	g_3(\vec{k}) &\equiv& 4\sqrt{3}\sin\left(\frac{ak_x}{2}\right)\sin\left(\frac{\sqrt{3}ak_y}{2}\right),\\
	g_4(\vec{k}) &\equiv& 4\sin\left(\frac{ak_x}{2}\right)\left[\cos\left(\frac{ak_x}{2}\right) - \cos\left(\frac{a\sqrt{3}k_y}{2}\right)\right],\nonumber
\end{eqnarray}

Expanding $g_2(\vec{k})$, $g_3(\vec{k})$ and $g_4(\vec{k})$ around Dirac's point up to first order in $\vec{p}$, we get
\begin{eqnarray}
	g_2(\vec{k}) &\sim& \xi \frac{3\sqrt{3}a}{\hbar}p_x, \nonumber\\
	g_3(\vec{k}) &\sim& \xi \frac{3\sqrt{3}a}{\hbar}p_y,\\
	g_4(\vec{k}) &\sim& -\xi 3\sqrt{3}.\nonumber
\end{eqnarray}

Terms $g_2(\vec{k})$ y $g_3(\vec{k})$ are negligible in front of $g_4(\vec{k})$ at zero order in $p$, therefore,
\begin{eqnarray}
	H_{AA}^{SO} &\sim& - \xi 3\sqrt{3}\Delta_{xz, yz}^{SO}s_z = - \xi \Delta_{xz, yz}^{SO}s_z,
\end{eqnarray}
where we redefine $\Delta_{xz, yz}^{SO}$ to absorb additional constant terms.

\section{Non-Diagonal Elements: Kinetic and SO Contributions of the Bloch Hamiltonian}\label{ndiagelem}
\subsubsection{Kinetic Contribution}
{In this appendix we show calculations for $H_{\rm AB}$ analogous to that of $H_{\rm BA}$ that leads to the complex conjugate of these results. Then, the non diagonal terms are calculated from }
\begin{eqnarray}
	H_{AB}^K 	&=& t_1\left[e^{i\vec{k}\cdot\vec{\delta}_1} + e^{i\vec{k}\cdot\vec{\delta}_2} + e^{i\vec{k}\cdot\vec{\delta}_3}\right] \nonumber\\
				& & + t_3\left[e^{i\vec{k}\cdot(\vec{\delta}_2 - \vec{\delta}_1 + \vec{\delta}_3)} + e^{i\vec{k}\cdot(\vec{\delta}_3 - \vec{\delta}_2 + \vec{\delta}_1)} \right.\nonumber\\
				& & \left.\qquad + e^{i\vec{k}\cdot(\vec{\delta}_1 - \vec{\delta}_3 + \vec{\delta}_2)}\right]\nonumber\\
				&=& t_1 f(\vec{k}) + t_3h_1(\vec{k}),
\end{eqnarray}
where
\begin{eqnarray}
	\begin{split}
		f(\vec{k}) &\equiv e^{i\frac{ak_y}{\sqrt{3}}} + 2e^{-i\frac{ak_y}{2\sqrt{3}}}\cos\left(\frac{ak_x}{2}\right), \\
		h_1(\vec{k}) &\equiv e^{-i\frac{2ak_y}{\sqrt{3}}} + 2e^{i\frac{ak_y}{\sqrt{3}}}\cos(ak_x).
	\end{split}
\end{eqnarray}

Expanding $f(\vec{k})$ and $h_1(\vec{k})$ around Dirac's point up to first order in $\vec{p}$, we get
\begin{eqnarray}
	\begin{split}
	f(\vec{k}) &\sim -\frac{\sqrt{3}a}{2\hbar}(\xi p_x - ip_y),\\
	h_1(\vec{k}) &\sim \frac{\sqrt{3}a}{\hbar}(\xi p_x - ip_y).
	\end{split}
\end{eqnarray}

Therefore,
\begin{eqnarray}
H_{AB}^{K} &\sim& \tilde{t}(\xi p_x - ip_y)
\end{eqnarray}
where
\begin{eqnarray}
\tilde{t} &\equiv& -\frac{\sqrt{3}a}{2\hbar}(t_1 - 2t_3) = v_F + \tilde{v}
\end{eqnarray}
with $v_F = -\sqrt{3}aV_{pp}^\pi / 2\hbar \approx 10^6 {\rm m/s}$ \cite{Castro} the Fermi velocity of the pristine graphene and 
\begin{equation}
\tilde v=-\frac{\sqrt{3}a}{2\hbar(\varepsilon_p-\varepsilon_d)}\left [\frac{V^2_{pd_{xz}}}{\zeta^2}-\frac{V^2_{pd_{xy}}}{12\zeta^2}\right ].
\end{equation}

\subsubsection{Spin-Orbit Contribution}

\begin{eqnarray}
	H_{AB}^{SO} &=& i\Delta_{xz,xy}^{SO}\left[-4s_xe^{i\vec{k}\cdot\vec{\delta}_1} + 2(s_x + \sqrt{3}s_y)e^{i\vec{k}\cdot\vec{\delta}_2} \right.\nonumber\\
				& & \left.\qquad\qquad + 2(s_x - \sqrt{3}s_y)e^{i\vec{k}\cdot\vec{\delta}_3}\right]\nonumber\\
				& & + i\Delta_{xz,xy}^{SO}\left[-2s_xe^{i\vec{k}\cdot(\vec{\delta}_2 - \vec{\delta}_1 + \vec{\delta}_3)} \right.\nonumber\\
				& & \qquad\qquad + (s_x + \sqrt{3}s_y)e^{i\vec{k}\cdot(\vec{\delta}_3 - \vec{\delta}_2 + \vec{\delta}_1)} \nonumber\\
				& & \left. \qquad\qquad + (s_x - \sqrt{3}s_y)e^{i\vec{k}\cdot(\vec{\delta}_1 - \vec{\delta}_3 + \vec{\delta}_2)}\right]\nonumber\\
				&=& i\Delta_{xz,xy}^{SO}[s_xh_2(\vec{k}) + s_yh_3(\vec{k})] \nonumber\\
				& & + i\Delta_{xz,xy}^{SO}[s_xh_4(\vec{k}) - is_yh_5(\vec{k})]
\end{eqnarray}
where
\begin{eqnarray}
	\begin{split}
		h_2(\vec{k}) &\equiv 2e^{-i\frac{a}{2}\left(k_x + \frac{k_y}{\sqrt{3}}\right)}\left[1 + e^{iak_x} - 2e^{i\frac{a}{2}(k_x + \sqrt{3}k_y)}\right], \\
		h_3(\vec{k}) &\equiv 2\sqrt{3}e^{-i\frac{a}{2}\left(k_x + \frac{k_y}{\sqrt{3}}\right)}\left(e^{iak_x} - 1\right),\\
		h_4(\vec{k}) &\equiv 2e^{-i\frac{2a}{\sqrt{3}}k_y}\left(e^{i\sqrt{3}ak_y}\cos(ak_x)- 1\right), \\
		h_5(\vec{k}) &\equiv 2\sqrt{3}e^{i\frac{a}{\sqrt{3}}k_y}\sin(ak_x).
	\end{split}
\end{eqnarray}

Expanding $h_2(\vec{k})$ , $h_3(\vec{k})$, $h_4(\vec{k})$ and $h_5(\vec{k})$ around Dirac's point up to zero order in $\vec{p}$, we get
\begin{eqnarray}
	\begin{split}
	h_2(\vec{k}) \sim -6, \quad h_3(\vec{k}) \sim i6\xi,  \\
	h_4(\vec{k}) \sim -3 , \quad h_5(\vec{k}) \sim -3\xi. 
	\end{split}
\end{eqnarray}

Therefore,
\begin{eqnarray}
	H_{AB}^{SO} &\sim& i\Delta_{xz,xy}^{SO}(-6s_x + i\xi 6s_y) + i\Delta_{xz,xy}^{SO}(-3s_x + i\xi 3s_y)\nonumber\\
				&\sim& \Delta_{xz,xy}^{SO}(-is_x + \xi s_y),
\end{eqnarray}
where we redefine $\Delta_{xz,xy}^{SO}$ to absorb additional constant terms.


\bibliography{graphene_au}

\end{document}